\documentclass[a4paper,fleqn,usenatbib,useAMS,usedcolumn]{mnras}

\usepackage[T1]{fontenc}
\usepackage{ae,aecompl}
\usepackage[utf8]{inputenc}

\usepackage{graphicx} 
\DeclareGraphicsExtensions{.pdf,.png,.jpg,.ps}
\graphicspath{{./}}

\usepackage{xcolor}
\usepackage[fleqn]{amsmath} 
\usepackage{amssymb}  
\usepackage{gensymb}
\usepackage{hyperref}
\usepackage[nolist]{acronym}
\usepackage{ulem}
\usepackage{bm}
\usepackage{etoolbox}
\usepackage[inline]{enumitem}	

\makeatletter
\patchcmd\@combinedblfloats{\box\@outputbox}{\unvbox\@outputbox}{}{%
  \errmessage{\noexpand\@combinedblfloats could not be patched}%
}%
\makeatother

\acrodef{COMPAS}{
Compact Object Mergers: Population Astrophysics and Statistics}

\definecolor{mustard}{rgb}{1.0, 0.86, 0.35}
\definecolor{cyan(process)}{rgb}{0.0, 0.72, 0.92}
\definecolor{ochre}{rgb}{0.8, 0.47, 0.13}
\definecolor{linesOne}{rgb}{0, 0.4470, 0.7410}
\definecolor{linesTwo}{rgb}{0.8500, 0.3250, 0.0980}
\definecolor{linesThree}{rgb}{0.9290, 0.6940, 0.1250}
\definecolor{linesFour}{rgb}{0.4940, 0.1840, 0.5560}
\definecolor{linesFive}{rgb}{0.4660, 0.6740, 0.1880}

\newcommand{\lone}[1]{\textbf{\textcolor{linesOne}{#1}}}
\newcommand{\ltwo}[1]{\textbf{\textcolor{linesTwo}{#1}}}
\newcommand{\lthree}[1]{\textbf{\textcolor{linesThree}{#1}}}
\newcommand{\lfour}[1]{\textbf{\textcolor{linesFour}{#1}}}

\newcommand\Fiducial{\texttt{Fiducial}}

\newcommand\rate{\mathcal{R}}

\newcommand\bayesFactor{\mathcal{K}}

\acrodef{DNS}{double neutron star}
\acrodef{DCO}{double compact object}
\acrodef{NS}{neutron star}
\acrodef{BH}{black hole}
\acrodef{GRB}{gamma--ray burst}
\acrodef{RLOF}{Roche-lobe overflow}
\acrodef{CE}{common envelope}

\acrodef{SN}{supernova}
\acrodefplural{SN}[SNe]{supernovae}
\acrodef{ECSN}{electron-capture supernova}
\acrodefplural{ECSN}[ECSNe]{electron-capture supernovae}
\acrodef{USSN}{ultra-stripped supernova}
\acrodefplural{USSN}[USSNe]{ultra-stripped supernovae}
\acrodef{CCSN}{core-collapse supernova}
\acrodefplural{CCSN}[CCSNe]{core-collapse supernovae}

\acrodef{MS}{main-sequence}
\acrodef{HG}{Hertzsprung-gap}
\acrodef{CHeB}{core helium burning}
\acrodef{EAGB}{early asymptotic giant branch}
\acrodef{HeMS}{helium main-sequence}
\acrodef{HeHG}{helium Hertzsprung-gap}

\newcommand\hubbleTimeGyrs{14.03}

\newcommand\formationVarZero{11.34}
\newcommand\formationVarOne{24.04}
\newcommand\formationVarTwo{24.54}
\newcommand\formationVarThree{28.05}
\newcommand\formationVarFour{30.95}
\newcommand\formationVarFive{9.16}
\newcommand\formationVarSix{15.11}
\newcommand\formationVarSeven{13.53}
\newcommand\formationVarEight{16.30}
\newcommand\formationVarNine{9.08}
\newcommand\formationVarTen{5.26}
\newcommand\formationVarEleven{9.54}
\newcommand\formationVarTwelve{14.14}
\newcommand\formationVarThirteen{15.31}
\newcommand\formationVarFourteen{6.69}
\newcommand\formationVarFifthteen{28.05}
\newcommand\formationVarSixteen{10.22}
\newcommand\formationVarSeventeen{20.09}
\newcommand\formationVarEightteen{24.06}
\newcommand\formationVarNineteen{14.29}

\newcommand\bayesVarZero{-16.78}
\newcommand\bayesVarOne{0}
\newcommand\bayesVarTwo{-3.12}
\newcommand\bayesVarThree{3.03}
\newcommand\bayesVarFour{-2.50}
\newcommand\bayesVarFive{-3.08}
\newcommand\bayesVarSix{-1.05}
\newcommand\bayesVarSeven{-3.19}
\newcommand\bayesVarEight{-0.07}
\newcommand\bayesVarNine{0.02}
\newcommand\bayesVarTen{1.76}
\newcommand\bayesVarEleven{-1.97}
\newcommand\bayesVarTwelve{2.54}
\newcommand\bayesVarThirteen{0.27}
\newcommand\bayesVarFourteen{-3.34}
\newcommand\bayesVarFifthteen{-2.67}
\newcommand\bayesVarSixteen{-0.07}
\newcommand\bayesVarSeventeen{-3.23}
\newcommand\bayesVarEighteen{-2.22}
\newcommand\bayesVarNineteen{-0.16}

\newcommand\fractionVarZero{0.61}
\newcommand\fractionVarOne{0.73}
\newcommand\fractionVarTwo{0.94}
\newcommand\fractionVarThree{0.76}
\newcommand\fractionVarFour{0.87}
\newcommand\fractionVarFive{0.83}
\newcommand\fractionVarSix{0.77}
\newcommand\fractionVarSeven{0.81}
\newcommand\fractionVarEight{0.85}
\newcommand\fractionVarNine{0.84}
\newcommand\fractionVarTen{0.59}
\newcommand\fractionVarEleven{0.36}
\newcommand\fractionVarTwelve{0.77}
\newcommand\fractionVarThirteen{0.72}
\newcommand\fractionVarFourteen{0.21}
\newcommand\fractionVarFifthteen{0.94}
\newcommand\fractionVarSixteen{0.69}
\newcommand\fractionVarSeventeen{0.71}
\newcommand\fractionVarEightteen{0.72}
\newcommand\fractionVarNineteen{0.70}

\newcommand\minFormTimeMyrsFiducial{8.5}
\newcommand\maxFormTimeMyrsFiducial{41.6}
\newcommand\minCoalTimeyrsFiducial{900.0}

\newcommand\minDelayTimeMyrsFiducial{12.6}

\newcommand\minMassRatioFiducial{0.58}
\newcommand\qAboveEightyFiducial{90}
\newcommand\qAboveNinetyFiducial{50}
\newcommand\qAboveNinetyFiveFiducial{30}
\newcommand\minMassRatioDelayed{0.52}
\newcommand\qAboveEightyDelayed{80}
\newcommand\qAboveNinetyDelayed{55}
\newcommand\qAboveNinetyFiveDelayed{40}

\newcommand\qAboveEightyMuller{70}
\newcommand\qAboveNinetyMuller{40}
\newcommand\qAboveNinetyFiveMuller{20}

\newcommand\doubleCoreCEE{23}
\newcommand\doubleECSN{0.1}
\newcommand\ECSNUSSN{19}

\newcommand\numberSecondariesUSSN{92}
\newcommand\numberECSN{20}

\newcommand\dominantFormationChannel{70}
\newcommand\secondFormationChannel{21}

\newcommand\minMassRapid{1.1}

\newcommand\minMassMuller{1.2}
\newcommand\maxMassRapid{1.9}

\newcommand\maxMassMuller{2.0}
\newcommand\heavierSecondaryAtZAMS{31}

\newcommand\cassBBsystemsAboveTwo{90}
\newcommand\cassBBsystemsAboveFour{9}

\newcommand\minSinglePrimaryECSN{7.8}
\newcommand\maxSinglePrimaryECSN{8.1}
\newcommand\minBinaryPrimaryECSN{7.8}
\newcommand\maxBinaryPrimaryECSN{28.4}
\newcommand\minSecondaryECSN{4.5}
\newcommand\maxSecondaryECSN{10.8}

\newcommand\typeIItoCCSN{75.6}
\newcommand\typeItoCCSN{24.4}
\newcommand\USSNtoCCSN{0.3}
\newcommand\USSNtoTypeI{1.2}
\newcommand\CCSNperMsolSF{0.0080}

\newcommand\evolvedMass{20,250,000}
\newcommand\totalMass{78,587,000}

\definecolor{indigo}{HTML}{FF00FF}
\renewcommand{\textsl}[1]{\textcolor{indigo}{\textbf{#1}}} 

\title[Double Neutron Stars]{On the formation history of Galactic double neutron stars
}

\author[]{\parbox{\textwidth}{
Alejandro Vigna-G\'{o}mez$^{1,2}$\thanks{E-mail: avigna@star.sr.bham.ac.uk},
Coenraad J. Neijssel$^{1}$,
Simon Stevenson$^{1,3}$,
Jim W. Barrett$^{1}$,
Krzysztof Belczynski$^{4}$,
Stephen Justham$^{5,6,2}$,
Selma E. de Mink$^{7}$,
Bernhard M\"uller$^8$,
Philipp Podsiadlowski$^{9,2}$,
Mathieu Renzo$^{7}$,
Dorottya Sz\'{e}csi$^{1}$,
Ilya Mandel$^{1,8,2}$
}
\vspace{0.5cm}\\
\parbox{\textwidth}{
$^{1}$ Birmingham Institute for Gravitational Wave Astronomy and School of Physics and Astronomy, University of Birmingham,\\
Birmingham, B15 2TT, United Kingdom \\
$^{2}$ DARK, Niels Bohr Institute, University of Copenhagen, Blegdamsvej 17, 2100, Copenhagen, Denmark \\
$^{3}$ OzGrav, Swinburne University of Technology, Hawthorn VIC 3122, Australia \\
$^{4}$ Nicolaus Copernicus Astronomical Center, Polish Academy of Sciences, ul. Bartycka 18, 00-716 Warsaw, Poland \\
$^{5}$ School of Astronomy \& Space Science, University of the Chinese Academy of Sciences, Beijing 100012, China \\
$^{6}$ National Astronomical Observatories, Chinese Academy of Sciences, Beijing 100012, China \\
$^{7}$ Anton Pannekoek Institute for Astronomy, University of Amsterdam, 1090 GE Amsterdam, The Netherlands \\
$^{8}$ Monash Centre for Astrophysics, School of Physics and Astronomy, Monash University, Clayton, Victoria 3800, Australia\\
$^{9}$ Department of Astronomy, Oxford University, Oxford OX1 3RH, UK
}
}

\date{\today}

\pubyear{2018}

\begin{document}

\label{firstpage}
\pagerange{\pageref{firstpage}--\pageref{lastpage}}
\maketitle

\begin{abstract}
Double neutron stars (DNSs) have been observed as Galactic radio pulsars, and the recent discovery of gravitational waves from the DNS merger GW170817 adds to the known DNS population. We perform rapid population synthesis of massive binary stars and discuss model predictions, including DNS formation rates, mass distributions, and delay time distributions. We vary assumptions and parameters of physical processes such as mass transfer stability criteria, supernova natal kick distributions, remnant mass prescriptions and common-envelope energetics. We compute the likelihood of observing the orbital period--eccentricity distribution of the Galactic DNS population under each of our population synthesis models, allowing us to quantitatively compare the models. We find that mass transfer from a stripped post-helium-burning secondary (case~BB) onto a neutron star is most likely dynamically stable. We also find that a natal kick distribution composed of both low (Maxwellian $\sigma=30\rm~km~s^{-1}$) and high ($\sigma=265\rm~km~s^{-1}$) components is preferred over a single high-kick component. We conclude that the observed DNS mass distribution can place strong constraints on model assumptions.  
\end{abstract}




\section{Introduction}
\label{sec:Introduction}
Since the first detection of a Galactic \ac{DNS} system \citep{hulse1975discovery}, the growing observed population of DNSs continues to provide constraints on their orbital parameters. Precise measurements of Keplerian and post-Keplerian parameters \citep{kramer2006tests} contain valuable information about the progenitors and formation history of \acp{NS} and \acp{DNS}.
Additionally, GW170817 \citep{GW170817} became the first gravitational--wave signal detected from a DNS merger. These precise measurements allow us to test our understanding on the massive binary progenitor populations and their formation history \citep[e.g.,][]{bhattacharya1991formation}.

\cite{tutukov1993formation} carried out an early rapid population synthesis study of Galactic NSs. The formation and fate of \acp{DNS} has been studied with a similar approach by \cite{portegies1998formation} who made an analysis of the observed systems and predictions of \ac{GRB} rates, and \cite{belczynski1999effect} who focused on gravitational--wave merger rates. \cite{voss2003mnras} studied both \ac{GRB} and gravitational--wave merger rates for Galactic \acp{DNS} (and binary black holes). \cite{Oshaughnessy:2005c} used six \acp{DNS} observed in the Galactic disk to constrain population synthesis models. Several binary population synthesis studies have focussed on natal kick distributions \citep[e.g.,][]{brandt1995dns,podsiadlowski2004effects,bray2016}, short \acp{GRB} locations \citep[e.g.,][]{church2011implications}, evolutionary channels \citep[e.g.,][]{andrews2015evolutionary} and merger rates \citep[e.g.,][]{chruslinska2018double}. More recently, \cite{2018arXiv180105433K} used their population synthesis model, calibrated to match the observed Galactic \ac{DNS} population, to predict merger rates in the local Universe.

Using the rapid population synthesis element of the Compact Object Mergers: Population Astrophysics and Statistics (COMPAS) suite \citep{stevenson2017formation}, we use the Galactic \ac{DNS} population as an observational constraint on massive binary evolution, from two zero age main sequence stars (ZAMS) to a pair of \acp{NS}. The COMPAS tool simulates isolated binaries: double star systems which evolve without significant interaction with the environment or with other stars.
The majority of the confirmed Galactic \acp{DNS} (14 confirmed systems, for details, see Table~\ref{tab:DNS}, as well as \cite{tauris2017formation} and references therein) come from isolated binaries which lie in the Galactic disk. We do not address the two Galactic globular cluster binaries in this work, $\rm~B2127+11C$ \citep{anderson1990discovery} and $\rm J1807-2500B$ \citep[][not a confirmed \ac{DNS}]{lynch2012timing}, since dynamical interactions likely played a key role in their formation \citep{phinney1991ejection,Grindlay2006sGRBs,ivanova2008NSGC}.

Our paper explores the impact of physical interactions during various stages of binary evolution on predictions of observables such as orbital parameters of Galactic \acp{DNS} and inferred mass distributions of gravitational--wave events. To do this, we compare models with different underlying assumptions and quantify the difference between their predictions. For each model, we provide \ac{DNS} formation rates and orbital parameters as predictions, to be tested against present time observations. We compare the predicted \ac{DNS} masses ($\rm m_{1,2}$) and orbital parameters (period P, eccentricity e) to those of the observed Galactic \acp{DNS}. We find that the natal kicks received by \acp{NS} during formation in a \ac{SN} and mass transfer stability criteria play a fundamental role in recreating the Galactic \ac{DNS} population. 

The paper is structured as follows. Section \ref{sec:methods} describes population synthesis and presents our \Fiducial~model. Changes made to binary evolution in COMPAS since \cite{stevenson2017formation} are specified. Section \ref{sec:results} presents the results of the \Fiducial~population, with particular emphasis on the formation history of Galactic \acp{DNS}. The effect of variations,  such as mass transfer during the post-helium-burning phase and the comparison between different natal kick distributions is mentioned. We conclude with a summary and discussion in Section \ref{sec:discussion}.

\begin{table}
\caption[caption]{Measured parameters of the Galactic \acp{DNS} used as a diagnosis in this study.
Notes: $^\dagger$Systems which will merge in via gravitational--wave emission in less than $3000~\rm Myrs$. $^{\ddagger}$Double pulsar. $^{\star}$Measurements used only for diagnosis in the $P-e$ plane. The masses of the \acp{DNS} are presented as $M_{\rm plsr}$ and $M_{\rm cmpn}$, the mass of the pulsar and the companion respectively.
References: $^a$\cite{martinez2015pulsar}. $^b$\cite{kramer2006tests}. $^c$\cite{fonseca2014comprehensive}. $^d$\cite{faulkner2004psr}. $^e$\cite{hulse1975discovery}. $^f$\cite{lazarus2016einstein}. $^g$\cite{cameron2017high}. $^h$\cite{janssen2008multi}. $^i$\cite{corongiu2007binary}. $^j$\cite{champion2004psr}. $^k$\cite{swiggum2015psr}. $^l$\cite{keith2009psr}. $^m$\cite{martinez2017pulsar}. $^n$\cite{stovall2018palfa}.} 
\label{tab:DNS}
\begin{tabular}{lcccccc}
\hline
Pulsar & $P$ & $e$ & $M_{\rm plsr}$  & $M_{\rm cmpn}$ & Ref \\
& $\rm [days]$ & & $\rm [M_{\odot}]$ & $\rm [M_{\odot}]$ &  \\
\hline
$\rm J0453+1559$ 			&	4.072	&	0.113	&	1.559	&	1.174	&	a	\\	
$\rm J0737-3039^{\dagger,\ddagger}$	& 	0.102	&	0.088	&	1.338	&	1.249	&	b	\\
$\rm B1534+12^{\dagger}$		&	0.421	&	0.274	&	1.333	&	1.346	&	c	\\
$\rm J1756-2251^{\dagger}$	&	0.320	&	0.181	&	1.341	&	1.230	&	d	\\
$\rm B1913+16^{\dagger}$		& 	0.323	&	0.617	&	1.440	&	1.389	&	e	\\
$\rm J1913+1102^{\dagger}$	& 	0.206	&	0.090	&	1.580	&	1.300	&	f	\\
$\rm J1757-1854^{\dagger}$	&	0.184	&	0.606	&	1.338	&	1.395	&	g	\\
$\rm J1518+4904^{\star}$	& 	8.634	&	0.249	&	-		&	-		&	h	\\
$\rm J1811-1736^{\star}$	& 	18.779	&	0.828	&	-		&	-		&	i	\\
$\rm J1829+2456^{\star}$	& 	1.176	&	0.139	&	-		&	-		&	j	\\
$\rm J1930-1852^{\star}$	& 	45.060	&	0.399	&	-		&	-		&	k	\\
$\rm J1753-2240^{\star}$	& 	13.638	&	0.304	&	-		&	-		&	l	\\
$\rm J1411+2551^{\star}$	& 	2.616	&	0.169	&	-		&	-		&	m	\\
$\rm J1946+2052^{\star}$	& 	0.078	&	0.064	&	-		&	-		&	n	\\
\end{tabular}
\end{table}

\section{Methods}
\label{sec:methods}

\subsection{Population Synthesis}
\label{subsec:pop_synth}
The COMPAS suite includes a rapid population synthesis module designed to simulate isolated binary evolution. Rapid population synthesis aims to simulate the evolution of a binary system in a fraction of a second; that makes it possible to simulate millions of binaries in a few days using a single processor. The population synthesis module of COMPAS explores binary evolution with a Monte Carlo simulation. We stochastically sample the initial distribution of binary masses, separations and eccentricities, in order to generate a population.

Given a certain mass and certain metallicity value at ZAMS, we define the initial conditions and evolution of a star following the fitting formulae of single-star evolution (SSE) as given in \cite{hurley2000comprehensive} to the detailed models calculated in \cite{pols1998stellar}. We use the same nomenclature as \cite{hurley2000comprehensive} to define stellar phases, such as the \ac{HG}, where the \ac{HG} is defined as the phase after the depletion of hydrogen during the \ac{MS} and before the start of \ac{CHeB}.
For every binary we follow the centre of mass evolution of the system, computing the masses, separation and eccentricity at every time step. We use parameterisations to quantify the effect on the orbit of the physics involving mass loss through stellar winds, mass transfer, \acp{SN} and \ac{CE} events. For \acp{SN} we also use remnant mass distributions which will determine the ultimate fate of our stars.
Each binary is evolved until the system either merges, becomes unbound or forms a \ac{DCO}. The population generates a set of \acp{DCO}, where \acp{DNS} are sub-selected into our final distribution of interest. COMPAS population synthesis is similar to the general approach of the codes SeBa \citep{portegies1996SEBA,portegies1998formation,toonen2012SEBA}, BSE \citep{hurley2002evolution}, \texttt{StarTrack} \citep{belczynski2002comprehensive,belczynski2008compact} and binary\_c \citep{izzard2004new,izzard2006population,izzard2009population}, all of which use the SSE fits from \cite{hurley2000comprehensive}.

Our current approach to the study of populations by proposing an initial model and studying the variations is similar to the one described in \cite{dominik2012double}. That study uses \texttt{StarTrack} to simulate populations from ZAMS to \ac{DCO} formation and predict merger rates for all compact objects. Their ``Standard" model overlaps with some of our \Fiducial~model assumptions.

\subsection{\Fiducial~Model}
\label{subsec:Fiducial}

\subsubsection[]{Changes since \cite{stevenson2017formation}}
\label{subsubsec:}
The main changes to binary evolution modelling in COMPAS relative to the default assumptions in \cite{stevenson2017formation}, hereafter referred to as $\rm COMPAS\_\alpha$, are:\\
\begin{enumerate*}[label=(\roman*), itemjoin={\\}, itemjoin*={{\\}}]
  \item incorporation of the fitting formulae of the binding energy parameter $\lambda_{\textrm{Nanjing}}$ instead of a fixed $\lambda=0.1$, as described in \ref{subsubsec:CE}. 
  \item a bimodal natal kick distribution, where \ac{CCSN} explosions contribute to the high mode ($\sigma_{\rm{high}} = 265~\rm km~s^{-1}$) while \ac{USSN} explosions and \ac{ECSN} explosions constitute the low mode ($\sigma_{\rm{low}} = 30~\rm km~s^{-1}$), as described in \ref{subsubsec:sne}.
  \item mass transfer stability criteria, allowing for always stable case~BB mass transfer, as described in \ref{subsubsec:CE}.
  \item the ``optimistic'' \ac{CE} assumption, which allows donors classified as Hertzsprung Gap (HG) stars in the \citet{hurley2000comprehensive} models  to engage and survive a \ac{CE} phase, as described in \ref{subsubsec:CE}.
  \end{enumerate*}

\subsubsection{Initial Distributions}
\label{subsubsec:initialDit}
To initialise a binary, we sample from initial distributions of mass, separation and eccentricity of the binary at ZAMS. For the mass distribution, we draw the primary mass from a Kroupa initial mass function (IMF) \citep{kroupa2001variation} in the form $dN/dm_1 \propto m_{1}^{-2.3}$ with masses between $5 \leq m_{1}/ \rm{M_{\odot}} \leq 100$. The secondary is drawn from a flat distribution in mass ratio between $0.1 < q_{\rm ZAMS} \equiv m_2/m_1 \leq 1$ \citep{sana2012binary}. The initial separation follows the flat-in-the-log distribution \citep{opik1924statistical,sana2012binary} in the range $0.1 < {a_{\rm ZAMS}}/ \rm{AU} < 1000.0$. We assume that all of our binaries are circular at ZAMS, with $e_{\rm ZAMS}=0$.

\subsubsection{Supernovae}
\label{subsubsec:sne}
We differentiate between three \ac{SN} scenarios: \ac{CCSN}, \ac{ECSN} and \ac{USSN}. For the \ac{CCSN} treatment, we apply the ``rapid" explosion mechanism\footnote{In this text, the term supernova explosion \textit{scenario} refers to the type of explosion, such as ECSN, USSN or CCSN, while the term explosion \textit{mechanism} refers to the numerical treatment of this process in the code. The latter is henceforth also referred to as supernova prescription, or fallback prescription, or remnant mass model.}, as presented in \cite{fryer2012compact}, to determine the compact object remnant mass according to the total and carbon-oxygen (CO) core mass of the progenitor, with a maximum allowed \ac{NS} mass of $m_{\rm{NS,max}} = 2.0~\rm M_{\odot}$. In this explosion mechanism, the collapse does not allow for accretion onto the proto-\ac{NS}, and is able to reproduce the apparent mass gap between \acp{NS} and black~holes \citep[BHs,][]{Ozel:2010,farr2011mass}. There is no consensus yet whether the mass gap is due to observational selection effects or if it is intrinsic to the explosion mechanism \citep{kreidberg2012mass,wyrzykowski2016black}. 

Another explosion scenario that some of our binary systems experience, is called \acp{USSN} \citep{tauris2013ultra,tauris2015ultra}. In this case, a star becomes stripped when it loses its hydrogen envelope during its evolution; if, during later stages, it manages to lose its helium envelope, it becomes ultra-stripped. In COMPAS, any star which engages in a stable case~BB mass transfer episode with a \ac{NS} as an accretor, is considered to be ultra-stripped. We define case~BB as a mass transfer episode which involves a post \ac{HeMS} donor star which has stopped burning helium in the core (helium~Hertzsprung-gap, HeHG). Ultra-stripped stars are left with an ONeMg core with a thin carbon and helium layer \citep{tauris2013ultra}. The compact object remnant mass of an \ac{USSN} is determined in the same way as for \ac{CCSN}.

A single star with mass within $7\lesssim$~$m_{\rm ZAMS}/\rm M_{\odot}\lesssim$~$10$ may collapse in an \ac{ECSN}. 
Early studies by \citet{nomoto1984evolution,nomoto1987evolution} had a higher mass range for single stars between $8\lesssim$~$m_{\rm ZAMS}/\rm M_{\odot}\lesssim$~$10$, while more recent studies propose a lower mass range from $7\lesssim$~$m_{\rm ZAMS}/\rm M_{\odot}\lesssim$~$9$ \citep{woosley2015ECSN}.
Note that binary interactions extend this initial mass range, which means that if we take a COMPAS simulation with binaries, the mass range for ECSNe will be broader because binarity changes the progenitor ZAMS masses (i.e. initially less massive stars which accreted mass or initially more massive which lost mass).
We assume the baryonic mass of the degenerate ONeMg core leading to an \ac{ECSN} is $1.38~\rm M_{\odot}$  \citep{nomoto1987evolution}. 
We approximate the \ac{ECSN} remnant mass as $m_{\rm{ECSN}}=1.26~\rm M_{\odot}$ using the quadratic approximation $m_{\rm{bar}}-m_{\rm{grav}}=0.075m_{\rm{grav}}^2$ \citep{timmes1995neutron}. 


All natal kicks from \acp{SN} are assumed to be isotropic in the frame of reference of the exploding star and randomly drawn from a unit sphere. For the magnitude of the natal kicks of the \acp{SN}, we assume a bimodal distribution \citep[e.g.,][]{1975Natur.253..698K,2002ApJ...568..289A}. For \ac{CCSN}, we draw natal kick magnitudes from a Maxwellian velocity distribution with a one-dimensional standard deviation of $\sigma_{\rm{high}} = 265~\rm km~s^{-1}$ following the 3D pulsar velocity distribution derived by \cite{hobbs2005statistical} from a subset of their 2D observations. \ac{USSN} and \ac{ECSN} natal kick magnitudes are drawn from a Maxwellian velocity distribution with a one-dimensional standard deviation of $\sigma_{\rm{low}} = 30~\rm km~s^{-1}$, following \cite{pfahl2002population} and \cite{podsiadlowski2004effects}. 
This second component is introduced to match the observed low natal kicks in some Galactic \acp{DNS} \citep{schwab2010further,beniamini2016formation} as well as in isolated pulsars \citep{0004-637X-571-2-906}, as the single-mode isolated pulsar velocity distribution proposed by \cite{hobbs2005statistical} fails to predict the low-velocity pulsar population as discussed by \cite{Verbunt2017bimodal} and \cite{verbunt2017new}.

\subsubsection{Mass Transfer}
\label{subsubsec:MT}
A crucial part of binary evolution is mass transfer, which begins when one or both stars fill their Roche lobe \citep{Eggleton1983rocheLobe}, instigating a \ac{RLOF} event. In our population synthesis approach, mass transfer is treated by determining stability, timescales and conservativeness.  Rapid population synthesis oversimplifies the complex  hydrodynamics involved in a mass transfer episode. There have been some efforts to provide generalised models \citep[e.g.,][]{deMink:2007SMC,claeys2014theoretical,tauris2015ultra}. In particular, determining whether mass transfer is dynamically stable is challenging \citep[e.g.,][]{pavlovskii2016stability}. 

To determine dynamical stability during mass transfer episodes, we compare the response of the donor star's radius to adiabatic mass loss, $\zeta_{\textrm{ad}}=(d\textrm{logR}/d\textrm{logM})_{\textrm{ad}}$, to the response of the Roche-lobe radius of the donor, $\zeta_{\textrm{RL}}$, under the same mass exchange conditions. Mass transfer is defined as dynamically stable if $\zeta_{\textrm{ad}}\geq\zeta_{\textrm{RL}}$. We use fixed values of $\zeta_{\textrm{ad,MS}}=2.0$ for \ac{MS} and $\zeta_{\textrm{ad,HG}}=6.5$ for \ac{HG} stars which are typical for these phases, following models by \cite{ge2015adiabatic}. For later phases in which the stars still possess hydrogen envelopes 
(such as the phases \ac{CHeB} and early asymptotic giant branch, EAGB)
we use a fit to $\zeta_{\textrm{ad}}=\zeta_{\textrm{SPH}}$ for condensed polytrope models of a red giant as provided in \cite{soberman1997stability}. Case BB mass transfer is always stable in the \Fiducial~model, broadly in agreement with \citet{tauris2015ultra}.

COMPAS uses fits to equilibrium mass-radius relations \citep{hurley2000comprehensive} to describe stellar evolution. We use these analytic formulae to determine when stable mass transfer is driven by thermal readjustment. If the calculated donor-star radius cannot stay within its Roche lobe during thermally stable mass transfer, then we remove the mass on a thermal timescale, although our stellar evolution recipes do not accurately represent the donor stars during thermal-timescale mass transfer \citep[for more detailed studies, see, e.g.,][]{kippenhahn1967entwicklung, pols1994caseA}. Once the donor's calculated equilibrium radius can again fit within its Roche lobe, we assume that the mass transfer occurs on a nuclear timescale \citep{claeys2014theoretical}.

Dynamically stable mass transfer from evolved stars is assumed to always proceed on the thermal timescale until the entire envelope is removed \citep[but see, e.g.,][]{2017A&A...608A..11G}. We approximate the thermal timescale as the Kelvin-Helmholtz timescale of the donor's envelope, $\tau_{\rm KH}=GMM_{\rm env}/RL$, where $G$ is the gravitational constant, $M$ is the total mass, $M_{\rm env}$ is the mass of the envelope, $R$ is the radius and $L$ is the luminosity of the star.

Conservativeness is defined as the amount of transferred mass from the donor that the accretor will accept and retain. When mass is lost from the system during non-conservative mass transfer, the fraction of mass lost and the specific angular momentum it carries away determine the orbital parameters and subsequent evolution of the system. In the \Fiducial~model, if mass transfer is non-conservative, the non-accreted mass is lost from the vicinity of the accreting star via isotropic re-emission, carrying away the specific orbital angular momentum of the accretor. The conservativeness of our mass transfer episode is limited by the accretor. For non-degenerate accretors we assume a star can accrete at a maximum rate $\dot{M}_{acc}=CM_{\rm acc}/\tau_{\rm KH}$. We use $C=10$ following \cite{hurley2002evolution}. For degenerate accretors, we assume the compact object accretion is limited by the Eddington accretion limit. 

\subsubsection{Common Envelope}
\label{subsubsec:CE}
If either of the binary stars begin dynamically unstable mass transfer, the binary may become engulfed in a \ac{CE} phase. The loss of corotation between the binary system and the envelope generates drag forces, which causes the binary to inspiral. The gravitational energy lost from the orbit is deposited in the envelope and may be enough to eject it from the binary. The whole process allows the system to decrease its separation several orders of magnitude.

The classical isolated binary evolutionary scenario for the formation of \acp{DCO} often involves a \ac{CE} phase \citep{paczynski1976common,ivanova2013common,belczynski2016first}.
We use the $\alpha\lambda$-formalism, as proposed by \cite{webbink1984double} and \cite{de1990common}, to estimate the effect of the \ac{CE} phase on the orbit of the binary.

The value of $\lambda$, which parametrises the envelope's binding energy, is calculated from detailed models of the stellar structure. For our \Fiducial~model, we adopt $\lambda_{\textrm{Nanjing}}$ (originally referred to as $\lambda_{\rm b}$, which includes internal energy) as calculated by \cite{xu2010binding}. This $\lambda_{\textrm{Nanjing}}$ is also implemented in the \texttt{StarTrack} code \citep{dominik2012double}.

The value of $\alpha$, which parametrises the efficiency of converting orbital energy into unbinding the envelope, depends on the orbital parameters, energy sources and energy exchange during the \ac{CE} phase, and is difficult to constrain even with detailed hydrodynamical models \citep{ivanova2013common}. We use $\alpha=1$. We assume that the orbit is always circularised during a \ac{CE} phase.   We allow donor stars which engage into a \ac{CE} phase during the \ac{HG} to survive the event and expel the \ac{CE} if allowed by the energy condition. This assumption is labeled ``optimistic'' \ac{CE} in the literature \citep{dominik2012double}, while the alternative, ``pessimistic'' \ac{CE}, always leads to a merger for \ac{HG} donors.

\subsection{Model Comparison}
\label{subsec:ModelComparison}

In order to quantify how well our models match the observed Galactic \ac{DNS} period--eccentricity ($P-e$) distribution, we calculate the likelihood $\mathcal{L}_i$ that observations could have come from the synthesised \ac{DNS} population for each model $i$. We use the $P-e$ distribution because all 14 observed Galactic \acp{DNS} (cf.~Table~\ref{tab:DNS}) have precise measurements of the period and the eccentricity, but only half of them have precise measurements of their individual masses. We do not use any of the mass measurements in the likelihood calculation.  We also do not attempt to account for selection biases in the observed $P-e$ distribution.	

\begin{figure}
\includegraphics[trim={0cm 0cm 1cm 4.5cm},clip,angle=90,scale=0.36]{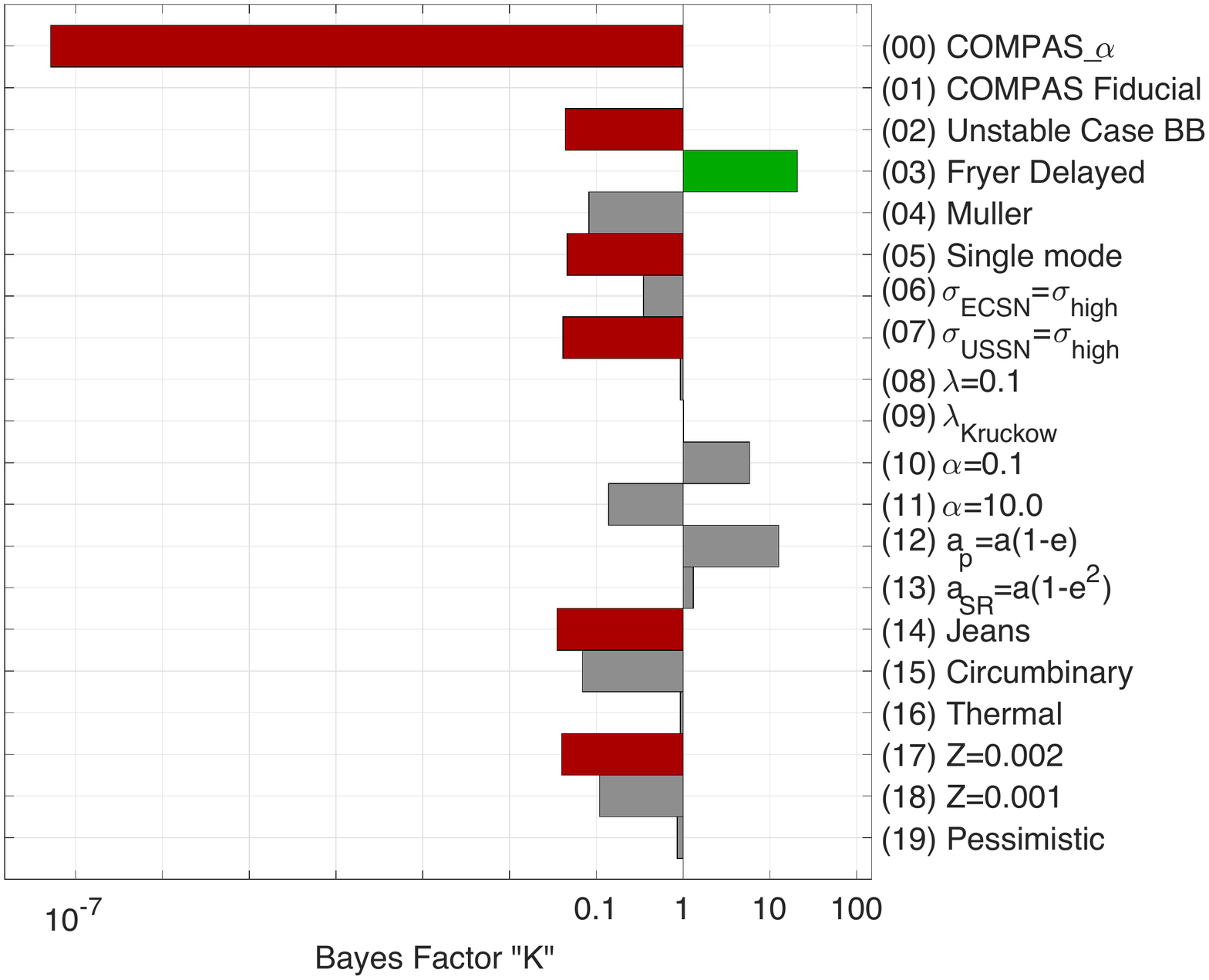}
\caption{The ratio of the likelihood of each model to the likelihood of the \Fiducial\ model (01). Green (red) bars denote models significantly favoured (disfavoured) by an odds ratio of greater than $20:1$ relative to the \Fiducial\ model.}\label{fig:likelihoods}
\end{figure}

The details of how the likelihoods $\mathcal{L}_i$ are computed are given in Appendix \ref{sec:likelihood}. We quote our results as the ratio of the likelihood for a given model to the likelihood of the \Fiducial~model $i$, that is, we define the Bayes~factor $\mathcal{K}_i$ as:
\begin{equation}
\log \mathcal{K}_i = \log \mathcal{L}_i - \log \mathcal{L}_{01} \, ,
\end{equation}
where all logarithms in this study are base $e$ unless stated otherwise. A positive log~Bayes~factor $\log~\mathcal{K}>0$ means that the given model is preferred over the \Fiducial~model. On the other hand, a negative log~Bayes~factor means that the \Fiducial~model is preferred over the given model. If all models have equal priori probabilities, the odds ratio is equal to the Bayes~factor. The odds ratio determines how significantly favoured or unfavoured the model is with respect to the \Fiducial~model; for an introduction to Bayesian analysis, see \cite{jaynes2003probability} and \cite{sivia1996data}. For readers unfamiliar with Bayes~factors, we also indicate when odds ratios for these model comparisons exceed $20:1$ (or $1:20$ for disfavoured models), loosely corresponding to the common significance threshold with a p--value of $p < 0.05$.  Limited sampling of the synthetic distributions leads to uncertainties of order unity on $\log \mathcal{K}_i$, corresponding to a factor of two or three uncertainty in the Bayes factor; this statistical uncertainty can be improved with longer simulations. The Bayes factors calculated for our models are plotted in Figure~\ref{fig:likelihoods} and presented in Table~\ref{tab:models}.  

\begin{table*}
\caption{
We list all simulations computed for this study; for simulations (02) through (19), we state the physical interaction or assumption varied relative to the \texttt{Fidicual} model and the actual parameter varied.
For each simulation, we give the formation rate $\rate$ of \ac{DNS} which will merge in a Hubble time in the Galaxy, its log~Bayes~factor relative to the \texttt{Fidicual} model (see Appendix \ref{sec:likelihood}) given the observed Galactic \ac{DNS} $P-e$ distribution and the fraction $f$ of formed \acp{DNS} that merge within a Hubble time.
See Figure~\ref{fig:nineteenpanels} for the predicted $P-e$ distributions for all models.
}
\label{tab:models}
\begin{tabular}{lcccccc}
\hline
Number & Physics & Variation  & $\rate$~[\textrm{$\rm Myr^{-1}$}]	& $\textrm{log}(\mathcal{K})$ & $f$ \\
\hline
00 & \texttt{$\rm COMPAS\_\alpha$} & & \formationVarZero & \bayesVarZero & \fractionVarZero \\
01 & $\rm COMPAS$ \texttt{Fiducial} & & \formationVarOne & \bayesVarOne & \fractionVarOne \\
02 & Stability & Case BB: unstable & \formationVarTwo & \bayesVarTwo & \fractionVarTwo \\
03 & SNe & Fryer Delayed & \formationVarThree & \bayesVarThree & \fractionVarThree \\
04 & SNe & M\"uller & \formationVarFour & \bayesVarFour & \fractionVarFour \\
05 & SNe & Single Mode & \formationVarFive & \bayesVarFive & \fractionVarFive \\
06 & SNe & $\sigma_{\textrm{ECSN}}=\sigma_{\textrm{high}}$ & \formationVarSix & \bayesVarSix & \fractionVarSix \\
07 & SNe & $\sigma_{\textrm{USSN}}=\sigma_{\textrm{high}}$ & \formationVarSeven & \bayesVarSeven & \fractionVarSeven \\
08 & CE & $\lambda=0.1$ & \formationVarEight & \bayesVarEight & \fractionVarEight \\
09 & CE & $\lambda_{\textrm{Kruckow}}\propto R^{-5/6}$ & \formationVarNine & \bayesVarNine & \fractionVarNine \\
10 & CE & $\alpha=0.1$ & \formationVarTen & \bayesVarTen & \fractionVarTen \\
11 & CE & $\alpha=10.0$ &\formationVarEleven & \bayesVarEleven & \fractionVarEleven \\
12 & Circularisation & $a_{\rm p}=a(1-e)$ & \formationVarTwelve & \bayesVarTwelve & \fractionVarTwelve \\
13 & Circularisation & $a_{\rm SR}=a(1-e^2)$ & \formationVarThirteen & \bayesVarThirteen & \fractionVarThirteen \\
14 & Mass Loss Mode & Jeans & \formationVarFourteen & \bayesVarFourteen & \fractionVarFourteen \\
15 & Mass Loss Mode & Circumbinary & \formationVarFifthteen & \bayesVarFifthteen & \fractionVarFifthteen \\
16 & Distribution & $f_{e}(e)=$ Thermal & \formationVarSixteen & \bayesVarSixteen & \fractionVarSixteen \\
17 & Metallicity & Z=0.002 & \formationVarSeventeen & \bayesVarSeventeen & \fractionVarSeventeen \\
18 & Metallicity & Z=0.001 & \formationVarEightteen & \bayesVarEighteen & \fractionVarEightteen \\
19 & CE & Pessimistic & \formationVarNineteen & \bayesVarNineteen & \fractionVarNineteen \\
\hline
\end{tabular}
\end{table*}

\section{Results}
\label{sec:results}
We evolve $10^6$ binaries\footnote{The total mass of evolved binaries is \evolvedMass $~\rm M_{\odot}$ for each simulation; this represents \totalMass $~\rm M_{\odot}$ of total star forming mass under the assumed initial mass distribution.} with initial metallicity $\rm Z_{\odot}=0.0142$. Although Galactic \acp{NS} were born at a range of metallicities, we use solar metallicity values \citep{asplund2009chemical} for bulk composition as a proxy for ongoing star formation metallicity in the Galaxy.

\begin{figure}
	\centering
	\includegraphics[trim={2.5cm 7.0cm 3.5cm 7.9cm},clip,scale=0.56]{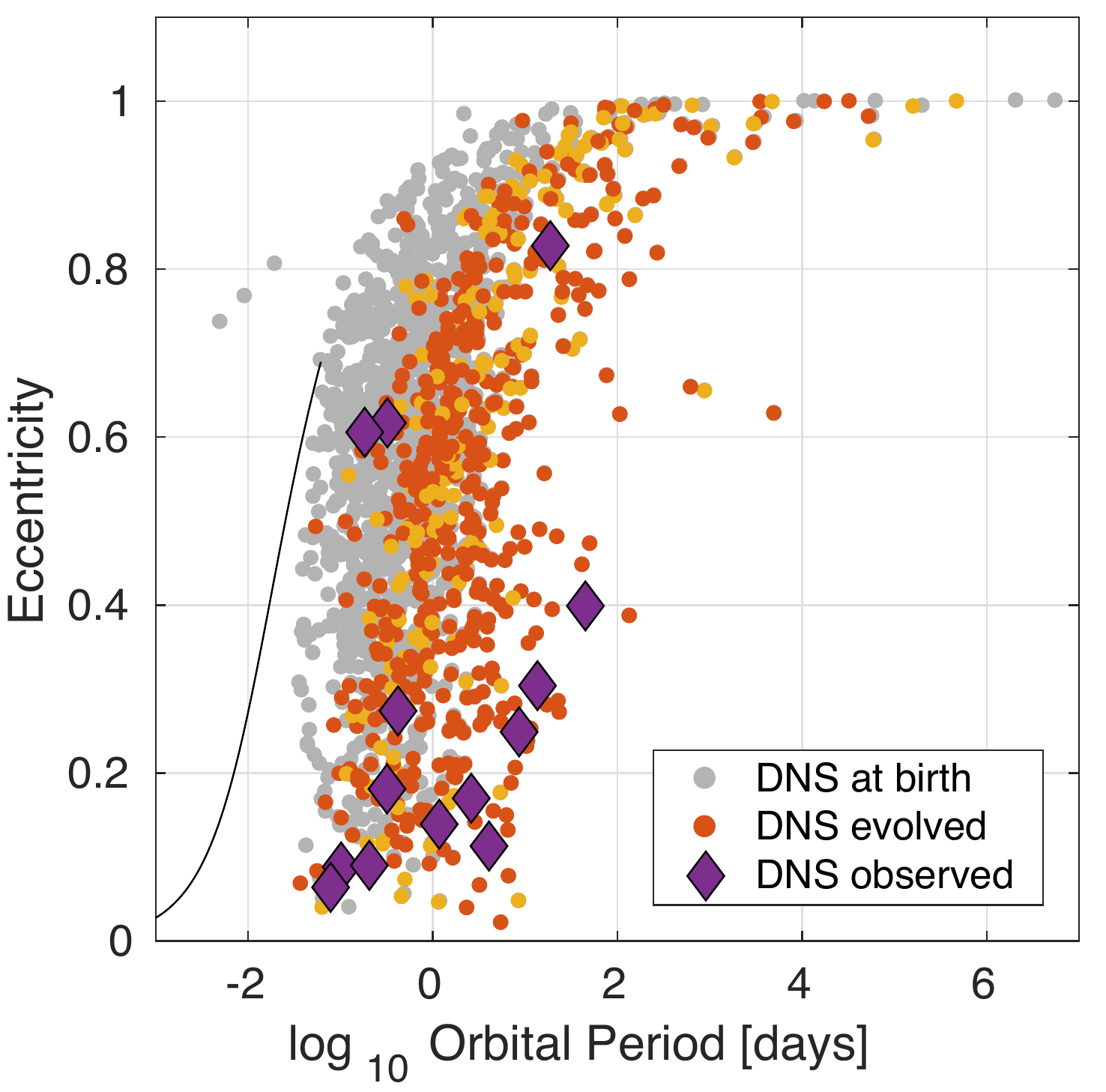}
	\caption{ Predicted $P-e$ distribution of Galactic \acp{DNS} under the \Fiducial~model. \textcolor{gray}{\textbf{Gray}} dots are all \acp{DNS} at \ac{DCO} formation. \ac{DCO} period and eccentricity are evolved forward from birth until present age given gravitational--wave radiation, removing a fraction of the short-lived short-period binaries from the observable population. Coloured dots represent the \ac{DNS} distribution at present age. Colour denotes the type of \ac{CE} phase: \ltwo{red} for a single-core, in which only the donor has a clear core-envelope separation, and \lthree{yellow} for a double-core \ac{CE} phase, in which both the donor and the accretor have a clear core-envelope separation. The single-core and double-core \ac{CE} phases can be, in most cases, associated with \textit{Channel~I} and \textit{Channel~II} respectively (see Section \ref{subsubsec:FormationChannels}, Figures \ref{fig:channelI} and \ref{fig:channelII} for more details). \textit{Channel~I} and \textit{Channel~II} are the first and second most common formation channels respectively. \lfour{Purple} diamonds represent the observed Galactic \acp{DNS}; all observed systems have precise $P-e$ measurements with error bars within the thickness of the symbol. The black curve illustrates a gravitational--wave driven $P-e$ evolution from \ac{DCO} formation to merger; this system,  with initial $P=1.5~\rm hours$, $e=0.69$ and characteristic \ac{NS} masses $m_{1}=m_{2}=1.2~\rm M_{\odot}$, would merge in $\approx 3$ Myr through gravitational--wave emission. }
	\label{fig:PeWG}
\end{figure}

We present the detailed results of our \Fiducial~model (01)\footnote{We will label the variations by their number (see Table \ref{tab:models}) in parentheses; e.g.: \Fiducial~model (01) or \texttt{COMPAS$_{\rm \alpha}$} (00).} and some variations to it, all with identical initial parameters (unless stated otherwise). The diagnostic tools we use to analyse all of our variations is the $P-e$ distribution (see Figure~\ref{fig:PeWG} and Section \ref{subsec:ModelComparison}, as well as Appendix \ref{sec:likelihood} for details), remnant \ac{NS} mass distribution (see Figure~\ref{fig:bnskde}) and formation rate estimates (see Table \ref{tab:models}). We report the number of significant figures based on statistical simulation uncertainty, i.e., the Monte Carlo uncertainty.

We illustrate the plausible distribution of simulated Galactic \acp{DNS} (see Figure~\ref{fig:PeWG} for \Fiducial~model and Figure~\ref{fig:nineteenpanels} for all models), which shows, in the $P-e$ plane, how systems may evolve from \ac{DNS} formation to a typical observable distribution. To illustrate this, we assign each binary a random probability of being born at any given point in the last 10 Gyr \citep[a proxy for the age of the Galactic thin disk, see][]{2005A&A...440.1153D}, and then follow their gravitational--wave driven orbital evolution until present time.

Our models predict the mass ratio distribution (Figure~\ref{fig:qCDFs}) and time distributions (Figure~\ref{fig:tdel}). The mass ratio distribution depends on the explosion mechanism of the \acp{SN}. The time distributions describe the formation time ($t_{\rm{form}}$), coalescence time ($t_{\rm c}$) and delay time ($t_{\rm{delay}}$). The formation time is the time it takes a binary to evolve from ZAMS to \ac{DCO} formation. The coalescence time is the time it takes that \ac{DCO} to inspiral until coalescence due to gravitational radiation, following the post-Newtonian approximation as given by \cite{peters1964gravitational}. The delay time is the sum of the formation time and the coalescence time.

Given the orbital properties of the population and the estimated time distributions we are able to predict the formation rate $\rate$ of \acp{DNS} which will merge in a Hubble time (assuming $H_0^{-1}=\hubbleTimeGyrs~\textrm{Gyr}$ in a flat $\rm\Lambda CDM$ cosmology; \citealt{ade2016planck}). If a system has a delay time of less than a Hubble time we include it in the formation rate $\rate$.

Formation rates are calculated for a galaxy with a continuous star formation rate of $f_{\textrm{SFR}}=2.0~\rm M_{\odot}/yr$ \citep{chomiuk2011toward}, with all systems in our simulated universe born in binaries\footnote{
While our models only include binaries, our orbital period distribution allows wide systems to evolve effectively as single stars. In fact, we find that more than half of our simulated binaries never engage in mass transfer.}. The star formation rate is chosen to mimic the Milky Way value of $f_{\textrm{SFR}}=1.9\pm0.4~\rm M_{\odot}/yr$ \citep{chomiuk2011toward}; any shifts in the chosen value would proportionately shift the quoted DNS formation rate.

A summary of all the formation rates and Bayes factors for the different variations is given in Table \ref{tab:models}. 

\subsection{On the \Fiducial~Model}
\label{subsec:FiducialResults}
\subsubsection{Formation Channels}
\label{subsubsec:FormationChannels}
There are two main ways that DNSs can form in our \Fiducial~model. We call these two dominant channels \textit{Channel~I} and~\textit{II}; some variations on these channels with additional mass transfer episodes or a different sequential order are possible.
Below we will explain the crucial steps in these formation channels, mentioning the fraction $f$ of systems that went through different stages of binary evolution. 
We find that 0.13 per cent of all simulated binaries become \acp{DNS}.

\textit{Channel~I}, illustrated in Figure~\ref{fig:channelI}, is responsible for the formation of roughly $\dominantFormationChannel$ per cent of all \acp{DNS}. This formation channel is consistent with the canonical channel described by, e.g., \cite{bhattacharya1991formation} and \cite{tauris2006formation}. \textit{Channel~I} involves a single-core \ac{CE} phase in which the primary has already collapsed into a \ac{NS}. A single-core \ac{CE} phase occurs when only one of the stars has a clear core--envelope separation; all compact objects are assumed not to have a clear core--envelope separation, as well as \ac{MS} and \ac{HeMS} stars. This channel proceeds as follows:\\

\textit{Channel~I}:\\
\begin{enumerate*}[label=(\roman*), itemjoin={{\\}}, itemjoin*={{\\}}]
  \item The two stars begin their evolution with the more massive one (the primary) evolving faster than its companion.
  \item \textit{$\approx$22 per cent of the all the initial systems experience stable mass transfer from the primary during the \ac{HG} phase onto a \ac{MS} secondary.}
  This is because $52$ per cent of the primaries never expand enough to start the mass transfer, and of the ones that do $47$ per cent are stable during this phase (0.48~$\times$~0.47~$\approx$~0.22).
  \item
  \textit{$\approx$4 per cent of those $\approx$22 per cent systems contain a primary that experiences a \ac{SN} explosion producing a \ac{NS} and remaining in a bound orbit.}
  In the mass transfer episode the primary becomes a \ac{HeMS} star. The majority of the \ac{HeMS} stars are either too light to become \acp{NS} or heavy enough to become BHs. Only 30 per cent of them have the mass of a \ac{NS} progenitor. In this first \ac{SN} explosion, there are ten times more \acp{CCSN} than there are \acp{ECSN} but, given the higher natal kick magnitude, their survival rate is only 9 per cent compared to 47 per cent of the \acp{ECSN}.
  \item
  \textit{$\approx$25 per cent of those $\approx$4 per cent systems experience and survive a \ac{CE} phase initiated by the post \ac{MS} secondary.} 
  Only 33 per cent of the secondaries expand enough to engage into a \ac{RLOF} mass transfer. This second mass transfer episode, with a primary \ac{NS} accretor, is usually dynamically unstable and leads to a \ac{CE} phase. 85 per cent of these systems are able to successfully eject their envelope, hardening the binary by two or three orders of magnitude.
    \item
	\textit{$\approx$40 per cent of those $\approx$25 per cent systems begin a third mass transfer episode (case~BB) of a \ac{HeHG} secondary onto a NS primary.}
	There the \ac{HeHG} star recycles its \ac{NS} companion while being stripped for a second time until a CO core (we call this ultra-stripped, see Sect.~\ref{subsubsec:sne}). Half of those cores are in the right mass range to become a \ac{NS} (lighter cores may form a \ac{NS}--white dwarf binary while heavier cores yield a \ac{NS}--BH binary).
  \item
  \textit
  {$\approx$96 per cent of those $\approx$40 per cent systems will remain bound after the second \ac{SN} explosion and form a \ac{DNS}.}
  The tight post--\ac{CE} orbit and the reduced natal kicks for \acp{USSN} make it relatively easy for binaries to survive the natal kick and form a \ac{DNS} system. The systems that are disrupted either lost enough mass and/or had orbital velocities low enough that even the reduced \ac{USSN} natal kick disrupts the system.
  \end{enumerate*}
\\

The secondary formation \textit{Channel~II}, illustrated in Figure~\ref{fig:channelII}, is responsible for forming approximately $\secondFormationChannel$ per cent of \acp{DNS}; it is prevalent for systems with initial mass ratio $q_{\rm ZAMS}\approx1$ and, therefore, similar evolutionary timescales of both stars in the binary. This channel experiences a double-core \ac{CE} phase \citep{brown1995doubleCore,dewi2006double,hwang2015twin}, in which both of the stars have a clear core-envelope separation, before the first \ac{SN}. \textit{Channel~II} proceeds as follows:\\


\textit{Channel~II}:\\
\begin{enumerate*}[label=(\roman*), itemjoin={{\\}}, itemjoin*={{\\}}]
	\item Again, the two stars begin their evolution with the primary evolving faster than its companion.
	\item
	\textit{ $\approx 1$ per cent of the primaries start their first mass transfer episode as either a \ac{CHeB} or an EAGB star with a secondary that is a slightly less evolved \ac{HG} or a \ac{CHeB} star.}
	Almost all of these systems (90 per cent) initiate a double-core \ac{CE} phase during this mass transfer episode.
	\item
	\textit{$\approx$35 per cent of those $\approx 1$ per cent binaries can eject their envelopes.}
	Only a tiny fraction ($\approx$2 per cent) lose enough mass to become white dwarfs whereas the majority become two \ac{HeMS} stars in a tighter orbit.
	\item
	\textit{$\approx$87 per cent of those $\approx$35 per cent systems have primaries that can initiate a second mass transfer episode (case~BB)}.
	The primaries donate their helium envelope to the secondary \ac{HeMS} star. All these episodes are dynamically stable.
	\item
	\textit{$\approx$35 per cent of those $\approx$87 per cent systems are able to have a primary experience a \ac{SN} explosion producing a \ac{NS} and remaining in a bound orbit}.
	As in \textit{Channel~I}, the mass transfer episodes reduce the masses of the primary and only 63 per cent can experience a \ac{SN} explosion. They are all \acp{CCSN} and although the \ac{CE} phase leaves them in a tight orbit the higher natal kick magnitude still disrupts 45 per cent of these systems.
	\item
	\textit{$\approx$80 per cent of those $\approx$35 per cent systems begin a third mass transfer episode (case~BB) from the secondary to a \ac{NS} accretor}.
	This mass transfer episode onto the \ac{NS} is defined to always be stable and the secondary now becomes an ultra-stripped CO core.
	\item
	\textit{$\approx$55 per cent of those $\approx$80 per cent systems have secondaries which experience and survive a \ac{SN} explosion and become \acp{NS}}.
	71 per cent of the CO cores are massive enough to explode as a \ac{SN}, and given the previous episode of mass transfer they are all \acp{USSN}. The lower natal kicks and tighter orbits help to get a survival rate of 77 per cent, leaving a \ac{DNS} system behind.
   \end{enumerate*}
\\

All simulated DNS systems are shown in the $P-e$ distributions in Figure~\ref{fig:PeWG}, \ref{fig:PeMain} and Appendix \ref{sec:likelihood}.
Most of the DNS systems that survived a single-core CE phase come from \textit{Channel~I}, while most of those that survived a double-core CE phase come from \textit{Channel~II}. The rest of the \acp{DNS}, about 9 per cent of the total, come from more exotic or fortuitous channels, including non--recycled \acp{DNS} ($\leq1$ per cent of all Galactic-like \acp{DNS}). Non-recycled \ac{DNS} progenitors are systems which never had stable mass transfer onto a \ac{NS} \citep{1538-4357-550-2-L183}, which leads to spin up and recycling; all of them experienced \acp{CE} in our models, which we assume to be inefficient at spinning up the \ac{NS} and suppressing its magnetic field \citep{MacLeodRamirezRuiz:2015}.

We find that our \Fiducial~model has a formation rate of $\rate=\formationVarOne$ per Milky Way equivalent galaxy per Myr. All of our \acp{DNS} experience and survive at least one \ac{CE} phase, \doubleCoreCEE \  per cent of them in a double-core scenario. \\

\begin{figure*}
	\includegraphics[trim={6.7cm 2.0cm 6.5cm 2.0cm},clip,angle=90,width=\textwidth]{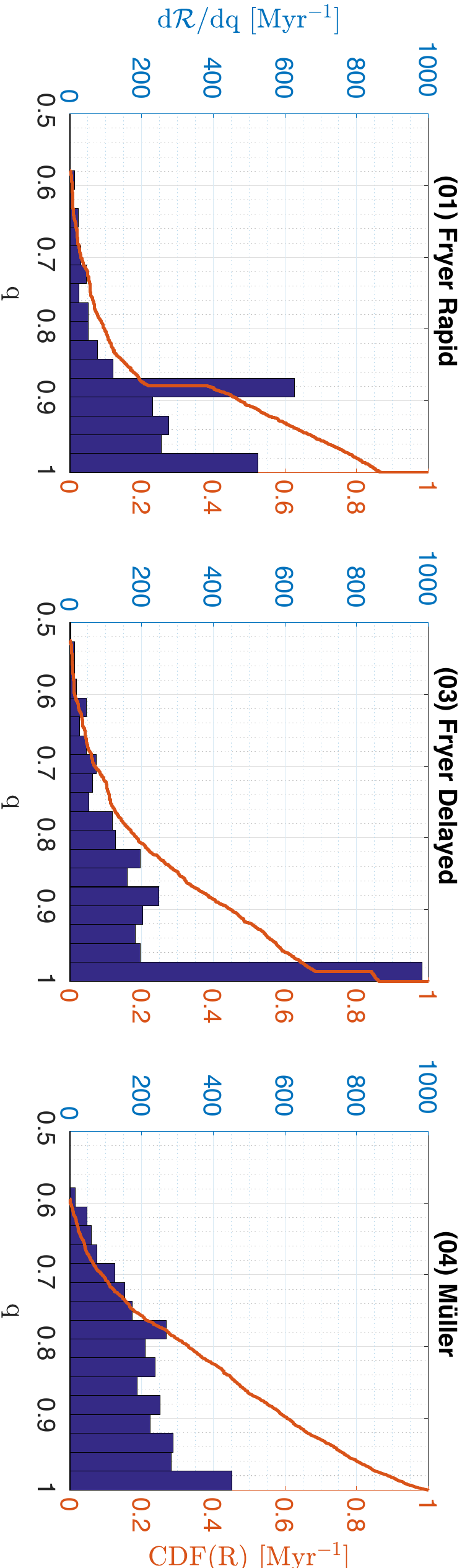}
	\caption[]{Mass ratio distribution of merging \acp{DNS} (blue histogram) and its cumulative distribution function (orange curve) for three \ac{SN} fallback and natal kick models: (01) Fryer Rapid [left], (03) Fryer Delayed [middle], (04) M\"uller [right]. See Sections \ref{subsubsec:FiducialMassRatio} and \ref{subsec:massRatio} for a discussion of the evolutionary channels leading to sharp features in the histograms.}
	\label{fig:qCDFs}
\end{figure*}

\begin{figure*}
	\includegraphics[trim={6.3cm 2.5cm 6.7cm 2.5cm},clip,angle=90,width=\textwidth]{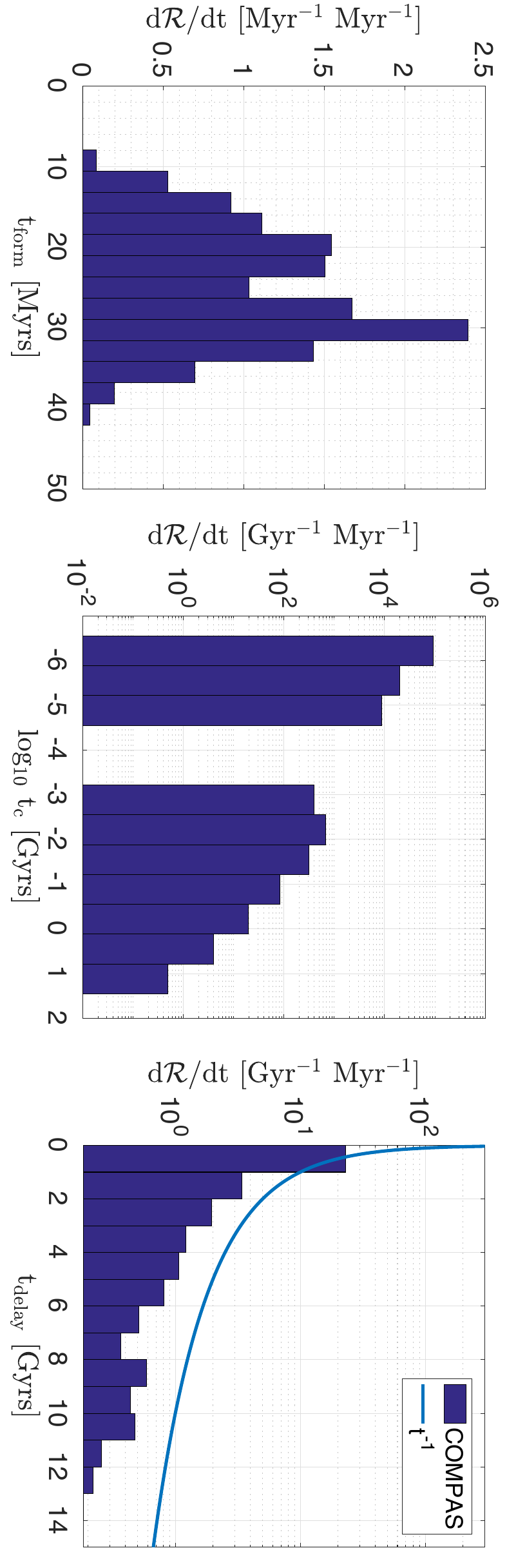}
	\caption{Time distributions or merging \acp{DNS} (blue histogram) for our \Fiducial~model (01): time $t_{\rm form}$ from ZAMS to \ac{DNS} formation [left], coalescence time $t_{\rm c}$ from \ac{DNS} formation to merger [middle] and total delay time $t_{\rm delay}$ from ZAMS to merger [right]. We show a $\rm d\rate/dt\propto t_{delay}^{-1}$ curve for comparison with the delay time distribution in the right panel. The apparent gap in the middle panel is a sampling artefact.} 
	\label{fig:tdel}
\end{figure*}

\begin{figure}
	\includegraphics[height=0.8\textheight]{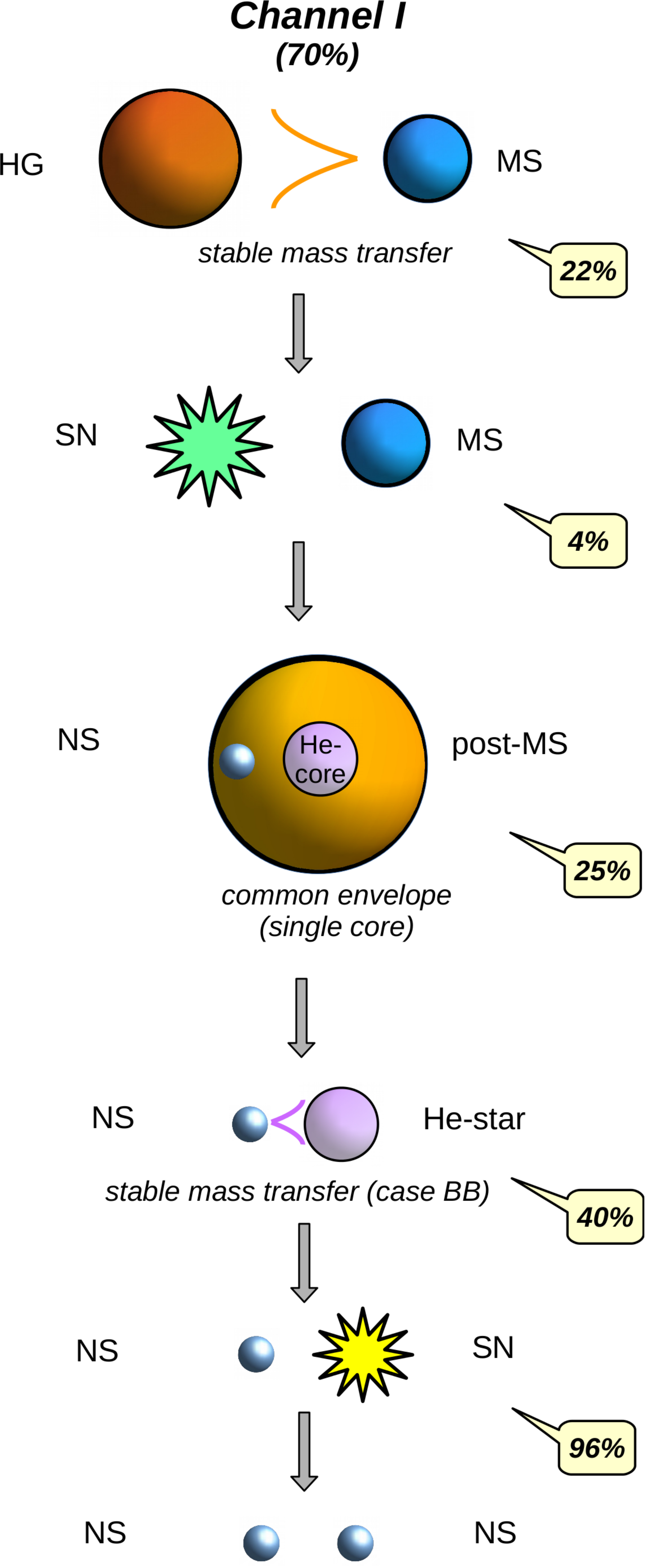}
	\caption{
	Evolutionary history of formation \textit{Channel~I} (top to bottom); 70 per cent of all \acp{DNS} in our \Fiducial~population were formed through this channel. The numbers in the callout symbols represent the percentage of simulated binaries that end up in that particular stage among those that follow the preceding evolutionary history. For example, $22$ per cent of all simulated binaries experience stable mass transfer from a \ac{HG} primary onto a \ac{MS} secondary; among those $22$ per cent, $4$ per cent of systems will have a primary that undergoes a \ac{SN} producing a \ac{NS} while remaining in a bound orbit; and so on.
	}
	\label{fig:channelI}
\end{figure}

\begin{figure}
	\includegraphics[height=0.8\textheight]{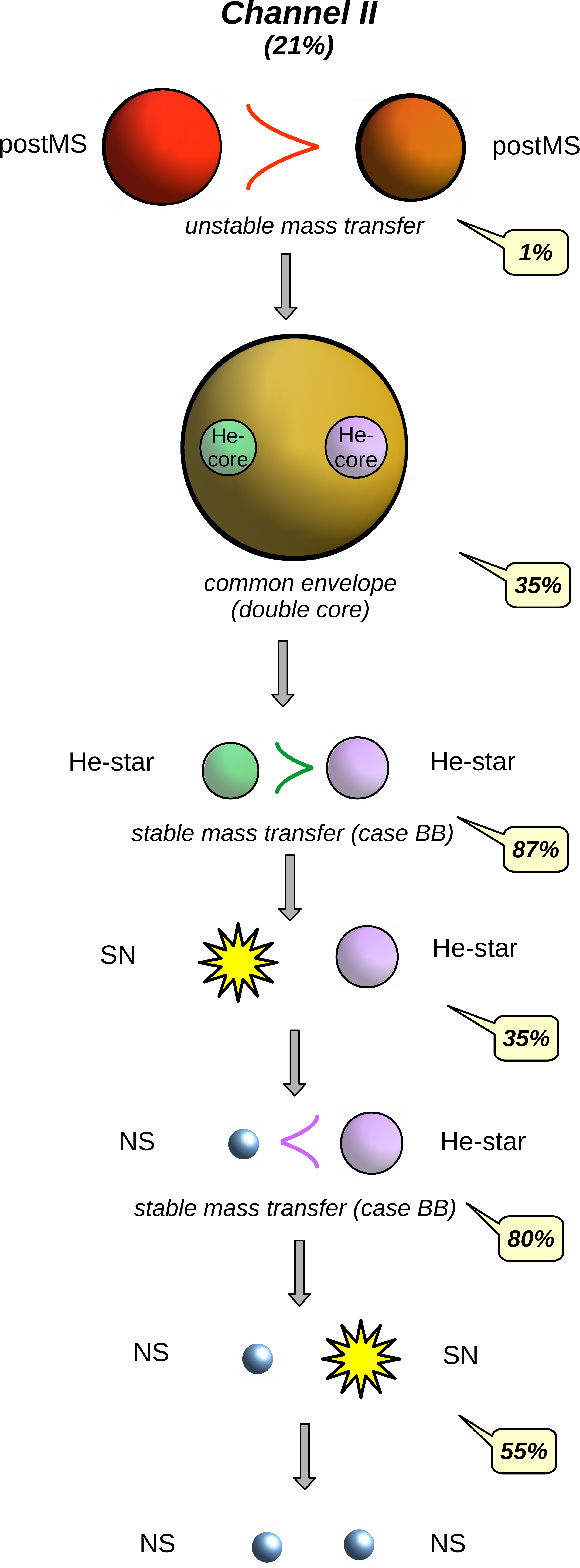}
	\caption{
	Evolutionary history of formation \textit{Channel~II} (top to bottom); 21 per cent of all \acp{DNS} in our \Fiducial~population were formed through this channel. The numbers in the callout symbols represent the percentage of simulated binaries that end up in that particular stage among those that follow the preceding evolutionary history.  For example, $1$ per cent of all simulated binaries initiate mass transfer while both companions are post-\ac{MS} stars; among those $1$ per cent, $35$ per cent enter and survive a double-core \ac{CE} phase; and so on.
	}
	\label{fig:channelII}
\end{figure}

\begin{figure*}
	\includegraphics[trim={2cm 1cm 2.0cm 2.5cm},clip,width=0.325\textwidth]{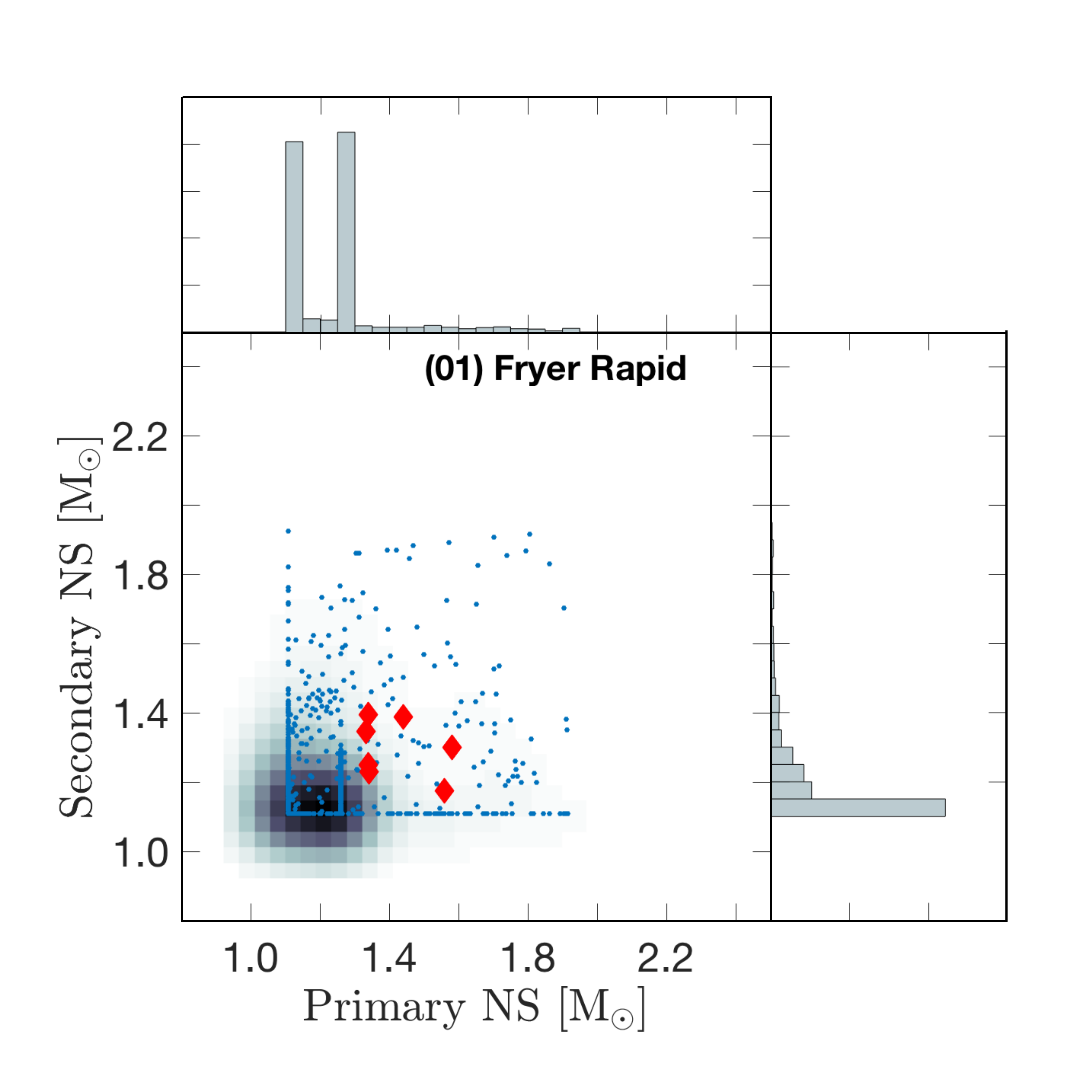}
	\includegraphics[trim={2.0cm 1cm 2.5cm 2.5cm},clip,width=0.325\textwidth]{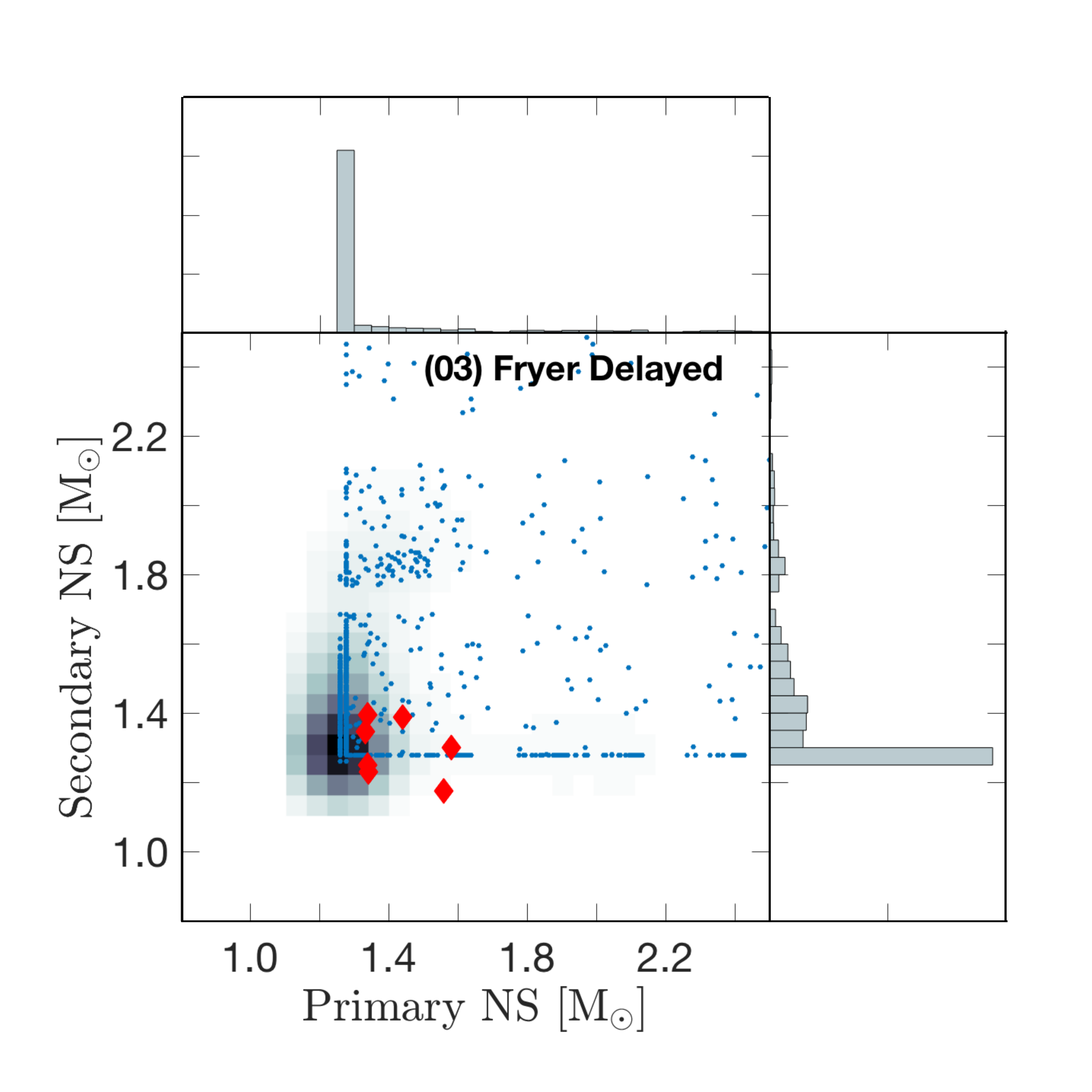}
	\includegraphics[trim={2cm 1cm 2.5cm 2.5cm},clip,width=0.325\textwidth]{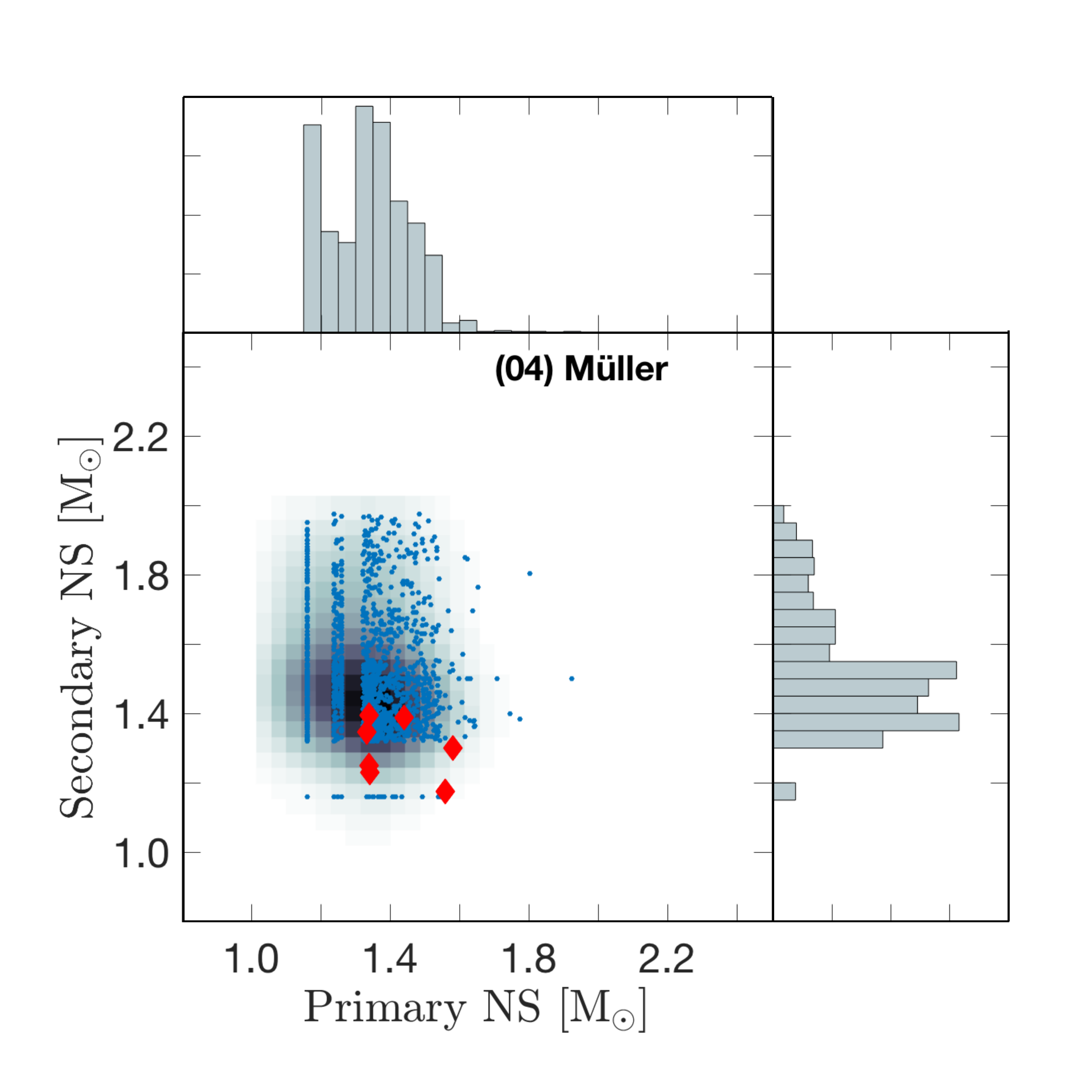}
	\caption{Predicted mass distribution of all predicted \acp{DNS} under three different \ac{SN} fallback and natal kick models: (01) Fryer Rapid [left], (03) Fryer Delayed [center], (04) M\"uller [right]. Primary and secondary mass of the \acp{NS} are shown in the horizontal and vertical axes respectively. \textcolor{red}{Red} diamonds denote the observed Galactic \acp{DNS} with well-constrained masses (see Table \ref{tab:DNS}), with pulsar and companion \ac{NS} masses shown in the horizontal and vertical axes respectively. \textcolor{blue}{Blue} dots correspond to the \ac{DNS} masses at \ac{DCO} formation. The density map shows the two-dimensional \ac{DNS} mass probability distribution; the histograms show its one-dimensional linear projections. See Sections \ref{subsubsec:FiducialMassRatio} and \ref{subsec:massRatio} for a discussion of the evolutionary channels leading to sharp features in the histograms.}
	\label{fig:bnskde}
\end{figure*}

\subsubsection{Mass Ratio Distribution}
\label{subsubsec:FiducialMassRatio}
Figure~\ref{fig:bnskde} shows the mass distribution of all the Galactic \acp{DNS} at the moment of birth, while Figure~\ref{fig:qCDFs} shows the distribution of the predicted mass ratio $q_{\rm DCO}$ for the merging Galactic \acp{DNS}. We define $q_{\rm DCO}=m_{\rm{NS,lighter}}/m_{\rm{NS,heavier}}$; the heavier \ac{NS} is not necessarily the more massive star at ZAMS. In the \Fiducial~model, the initially less massive star produces the more massive \ac{NS} in $\heavierSecondaryAtZAMS$ per cent of the systems, due to the accretion of mass from the companion, and its core growth, during the early phases of evolution.
The mass ratio of these systems lies between $\minMassRatioFiducial \leq q_{\rm DCO} \leq 1$. Among the merging Galactic \acp{DNS}, \qAboveEightyFiducial \  per cent of the systems have $q_{\rm DCO} > 0.8$, \qAboveNinetyFiducial \  per cent have $q_{\rm DCO} > 0.9$ and \qAboveNinetyFiveFiducial \ per cent have $q_{\rm DCO} > 0.95$. There are two significant peaks in this distribution: 
\begin{enumerate*}[label=(\roman*), itemjoin={{, }}, itemjoin*={{, and }}]
  \item the first peak, with $\approx16$ per cent of systems have $q_{\rm DCO}\approx0.88$; most systems close to this mass ratio are formed through \textit{Channel~I}, with their first \ac{NS} being an \ac{ECSN} (with gravitational mass of $1.26~\rm M_{\odot}$) and the second an \ac{USSN} (with lower mass remnants of \minMassRapid$~\rm M_{\odot}$)
  \item the second peak, with $\approx14$ per cent of the total \acp{DNS}, has a mass ratio $q_{\rm DCO}\approx1$, from $q_\mathrm{ZAMS} \approx 1$ systems that evolved through the double-core CE, with a low mass \ac{CCSN} and an \ac{USSN} (i.e. \textit{Channel~II}).
\end{enumerate*}
The mass range of NSs in our \Fiducial~population is $[m_{\rm{NS,min}}, m_{\rm{NS,max}}]=[\minMassRapid,\maxMassRapid]~\rm M_{\odot}$.

\subsubsection{Time Distributions}
\label{subsubsec:FiducialTimeDistribution}
We define the following timescales:
\begin{enumerate*}[label=(\roman*), itemjoin={{, }}, itemjoin*={{, and }}]
 \item formation time $t_{\rm form}$ as the time from ZAMS to \ac{DCO} formation 
 \item coalescence time $t_{\rm c}$ as the time from \ac{DCO} formation to merger
 \item total delay time $t_{\rm delay}$ as the time from ZAMS to merger.
\end{enumerate*}
Figure~\ref{fig:tdel} shows the distributions for $t_{\rm form}$, $t_{\rm c}$ and  $t_{\rm delay}$ for our \Fiducial~model. Time distributions were made based on only those \acp{DNS} which have a merger time of less than the Hubble time. 
The extreme ends of the time distributions are systems with $\minFormTimeMyrsFiducial \leq t_{\rm{form}}/\textrm{Myr} \leq \maxFormTimeMyrsFiducial$,  $\minCoalTimeyrsFiducial \leq t_{\rm{c}}/\textrm{yr}$  and $\minDelayTimeMyrsFiducial \leq t_{\rm{delay}}/\textrm{Myr}$. 
 
Fewer than 0.5 per cent of merging \acp{DNS} have very short coalescence times of less than 10 Myr (see middle panel of Figure~\ref{fig:tdel} and outliers in Figure~\ref{fig:kicks}---note that the apparent gap in the middle panel is a sampling artefact, and does not represent an actual gap in the population). These systems usually experience \acp{CE}, reduce their orbit during case~BB mass transfer and have fortuitous natal kick directions which place them on a low-periapsis orbit at \ac{DCO} formation. Systems with $t_{\rm{c}}>10^{-3}$~Gyr represent the bulk of the population in Figure~\ref{fig:PeWG}; shorter coalescence times are exhibited by outliers with orbital periods of $\lesssim 10^{-2}$~days.

\subsubsection{Supernovae}
\label{subsubsec:FiducialSupernovae}
Of all the \acp{NS} in \ac{DNS} systems, \numberECSN~per cent were formed via \acp{ECSN}. \numberSecondariesUSSN~per cent of the initially less massive secondaries in these \acp{DNS} experienced ultra-stripping before exploding.
Only \doubleECSN~per cent of the \ac{DNS} systems had both stars experiencing a \ac{ECSN}.
In \ECSNUSSN~per cent of the \acp{DNS} the primary went through an \ac{ECSN} and was later recycled by case BB mass transfer from the secondary.  The secondary is stripped by this mass transfer and explodes in an \ac{USSN}.


In our single stellar models at $Z=Z_{\rm \odot}$, \ac{ECSN} progenitors have masses at ZAMS of $\minSinglePrimaryECSN \leq m/\rm{M_{\odot}} \leq \maxSinglePrimaryECSN$; more recent detailed models find that the mass range of single star progenitors at metallicity $Z=0.02$ which explode as an \ac{ECSN} is $7.5 \leq m/\rm{M_{\odot}} \leq 9.25$ \citep{poelarends2008}. Interaction during binary evolution increases this range to $\minBinaryPrimaryECSN \leq m_1/\rm{M_{\odot}} \leq \maxBinaryPrimaryECSN$ for the primary and $\minSecondaryECSN \leq m_2/\rm{M_{\odot}} \leq \maxSecondaryECSN$ for the secondary in our study. Detailed studies of \acp{ECSN} from interacting binary systems find that the mass range for an interacting primary at $Z=0.02$ is between $13.5 \leq m/\rm{M_{\odot}} \leq 17.6$ \citep{poelarends2017}, where $17.6~\rm M_{\odot}$ is the highest mass primary used in that study.

Metallicity does not play a strong role in modifying the \ac{NS} mass range. We explore lower metallicity populations at $Z=0.002$ (17) and $Z=0.001$ (18), and find that, for single stars, the \ac{ECSN} progenitor masses at ZAMS decrease to $7.0 \leq m/\rm{M_{\odot}} \leq 7.2$ and $6.6 \leq m/\rm{M_{\odot}} \leq 6.9$, respectively. However, the remnant mass from an \ac{ECSN} does not vary as a function of metallicity and is always fixed to be $m_{\rm{ECSN}}=1.26~\rm M_{\odot}$. Furthermore, our minimum and maximum \ac{NS} masses of $[m_{\rm{NS,min}}, m_{\rm{NS,max}}]=[\minMassRapid,\maxMassRapid]~\rm M_{\odot}$ do not change as a function of metallicity.

\subsection{Variations}
COMPAS is a modular code designed to explore the effects of different treatments of uncertain physical assumptions. Given the complexity of the formation channels we explore the uncertainties by changing one assumption per variation. This allows us to link all the changes in the population and its formation channels to a specific physical treatment and test the robustness of our \Fiducial~model. The parameters of the physical interactions may be correlated; since computing these correlations is computationally expensive \citep[see e.g.][]{barrett2017accuracy} we do not consider them here.

\subsubsection{On Mass Transfer Stability Criteria}
\label{subsubsec:MTstability}
\begin{figure*}
	\centering
		\includegraphics[trim={7.1cm 1.0cm 6.9cm 2.8cm},clip,angle=90,scale=0.7]{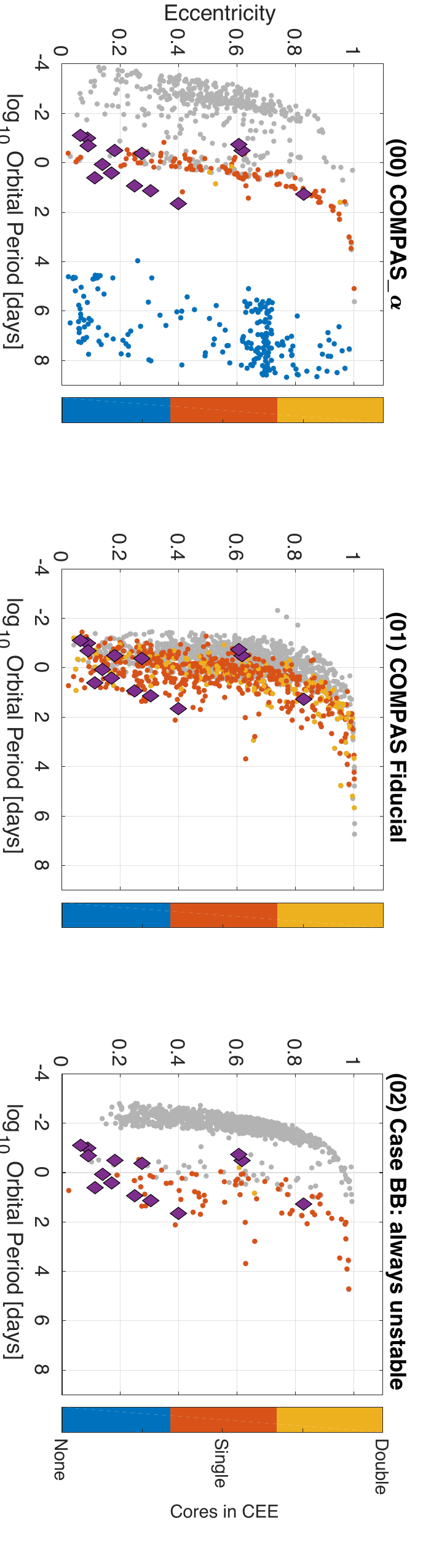}
	\caption[]{Predicted $P-e$ distribution of Galactic \acp{DNS} at \ac{DCO} formation: (00) \cite{stevenson2017formation} standard [left], (01) \Fiducial~[middle], (02) variation with unstable case~BB mass transfer [right] (for more details see Table \ref{tab:DNS}). \lfour{Purple} diamonds represent the Galactic \acp{DNS}. Colour denotes the type of \ac{CE} phase: \lone{blue} for no \ac{CE} phase, \ltwo{red} for a single-core and \lthree{yellow} for a double-core \ac{CE} phase. The single-core and double-core \ac{CE} formation are typically associated with \textit{Channel~I} and \textit{Channel~II}, respectively. \lone{Blue} dots on the left panel correspond to double-\acp{ECSN} with $\sigma_{\rm ECSN}=0\rm~km~s^{-1}$ in $\rm COMPAS\_\alpha$.} 
	\label{fig:PeMain}
\end{figure*}

Stable case~BB mass transfer leads to orbital periods similar to the observed Galactic \ac{DNS} population. Meanwhile, unstable case~BB, leading to a \ac{CE} phase, typically results in sub-hour orbital periods (see right panel of Figure~\ref{fig:PeMain}); such orbital periods yield coalescence times of $\lesssim 10$~Myr. About $~90$ per cent of Galactic \ac{DNS} progenitors in the \Fiducial~model experience case~BB mass transfer. At the onset of the episode, \cassBBsystemsAboveTwo \  per cent of systems have mass ratio $q \geq 0.2$ and \cassBBsystemsAboveFour \  per cent with $q \geq 0.4$. \cite{claeys2014theoretical} assume that mass transfer of \ac{HeHG} donors with a degenerate accretor will be stable if $q > 0.21$ (see Table 2 of that paper), while \cite{tauris2015ultra} propose to consider mass ratio and orbital period to define stability criteria in order to account for the evolutionary phase of the donor at the onset of \ac{RLOF}; in that study, orbital periods of $P \geq 0.07\rm~days$ at the onset of \ac{RLOF} lead to stable case~BB mass transfer. In our \Fiducial~model, all Galactic \ac{DNS} progenitors have $P \geq 0.07\rm~days$ at the onset of case~BB mass transfer. 

In $\rm COMPAS$, we probe the extreme cases of either stable or dynamically unstable case~BB mass transfer for a whole population. The difference in formation rate $\rate$ between the stable (01) and dynamically unstable (02) case~BB mass transfer is comparable within a few per cent, with $\{\rate_{\rm01}$,$\rate_{\rm02}\}=\{{\formationVarOne,\formationVarTwo}\}$ per Galaxy per Myr. Nevertheless, the log~Bayes~factor of model (02) relative to model (01) is log $\bayesFactor=\bayesVarTwo$, which favours our \Fiducial~model, and ultimately,  significantly favours stable against unstable mass transfer in a dichotomous scenario.
In our \Fiducial~population, the assumption of case~BB mass transfer being always stable is in broad agreement with mass ratio constraints from \cite{claeys2014theoretical}, which would result in more than 90 per cent of these systems experiencing stable mass transfer. If instead we used the the stability criteria presented in \cite{tauris2015ultra}  \citep[as shown in][]{2018arXiv180105433K}, all of the aforementioned systems would experience stable mass transfer.

\subsubsection{On the ``Delayed" Explosion Scenario}
To test the effect of the explosion mechanism on our predictions, we investigate three prescriptions; one being the ``rapid'' (01) explosion mechanism as presented in our \Fiducial~model (see Sect.~\ref{subsubsec:FiducialSupernovae}). The second one is the ``delayed'' (03) explosion mechanism applied in our model (03) and to be explained below, while the third is the ``M\"uller'' (04) prescription presented in Sect.~\ref{sec:Muller}.

The ``delayed'' explosion mechanism proposed in \cite{fryer2012compact} allows for accretion onto the proto-\ac{NS} before the standing-accretion shock instability or convection become powerful enough to support a neutrino-driven explosion. This accretion removes the mass gap and creates a continuous remnant mass distribution from \ac{NS} to BH. This continuous distribution of compact-object remnant masses requires us to define an arbitrary mass cut to distinguish \acp{NS} from BHs; we follow \cite{fryer2012compact} and set this mass cut to $2.5\rm~M_\odot$, which is higher than the maximum mass of $2.0\rm~M_\odot$ from the other explosion mechanisms used in this study.

The ``delayed'' explosion mechanism formation rate is $\rate=\formationVarThree$ per Milky Way equivalent galaxy per Myr.  The ``delayed" explosion mechanism, which changes the remnant mass given a CO core at the moment of a \ac{SN}, produces a slightly different $P-e$ distribution than the ``rapid'' explosion mechanism because of the impact of mass loss at the moment of the explosion on the binary's orbit. Middle panel of Figure~\ref{fig:bnskde} shows that the ``delayed" explosion mechanism lies close to the observed population and is preferred over the ``rapid'' explosions mechanism in the \Fiducial~model with a log~$\bayesFactor=\bayesVarThree$. The ``delayed" explosion scenario, which does not have a mass gap between \acp{NS} and BHs, has the largest likelihood of all models.

\subsubsection{On the Supernovae Kick Distribution and Magnitude}
Both mass loss during a \ac{SN} and the natal kick magnitude and direction modify the orbital parameters and determine whether the binary is disrupted. Low natal kick \acp{ECSN} and \acp{USSN} therefore play a prominent role in \ac{DNS} formation and possible eventual merger, as would low-mass iron-core-collapse \acp{SN} with a reduced natal kick. Our modelling allows for testing a bimodal natal kick distribution, which distinguishes between \acp{CCSN} (high mode, $\sigma_{\rm{high}} = 265~\rm km~s^{-1}$), \acp{ECSN} (low mode, $\sigma_{\rm{low}} = 30~\rm km~s^{-1}$) and \acp{USSN} (low mode). When allowing for a bimodal distribution, but with only \ac{USSN} (06) or \ac{ECSN} (07) contributing to the low component of the Maxwellian distribution, the \ac{DNS} formation rate $\rate$ drops by a factor of $\approx 2$ relative to the \Fiducial~model. We also simulated a single high-mode distribution (05) with high natal kicks for both \acp{USSN} and \acp{ECSN}, which is also the assumption in $\rm COMPAS\_\alpha$ (00). In this case, $\rate$ decreases by a factor of $\approx 3$; this single high-mode variation (05) also fails to create the observed longer period \acp{DNS} with low eccentricities. The formation rates and log~Bayes~factors are $\{\rate_{\textrm{05}},\rate_{\textrm{06}},\rate_{\textrm{07}}\}=\{\formationVarFive,\formationVarSix,\formationVarSeven\}$ per Milky Way equivalent galaxy per Myr and $\log~\{\bayesFactor_{\textrm{05}},\bayesFactor_{\textrm{06}},\bayesFactor_{\textrm{07}}\}=\{\bayesVarFive,\bayesVarSix,\bayesVarSeven\}$ for variations with a single high mode (05), $\sigma_{\rm ECSN}=\sigma_{\rm high}$ (06) and $\sigma_{\rm USSN}=\sigma_{\rm high}$ (07), respectively. Given the log~Bayes~factors, the \Fiducial~model is significantly preferred over single high mode (05) and $\sigma_{\rm USSN}=\sigma_{\rm high}$ (07) variations. It is preferred, but not significantly, over the $\sigma_{\rm ECSN}=\sigma_{\rm high}$ (06) variation.

\begin{figure*}
	\includegraphics[trim={7cm 0cm 7cm 0cm},clip,width=\textwidth]{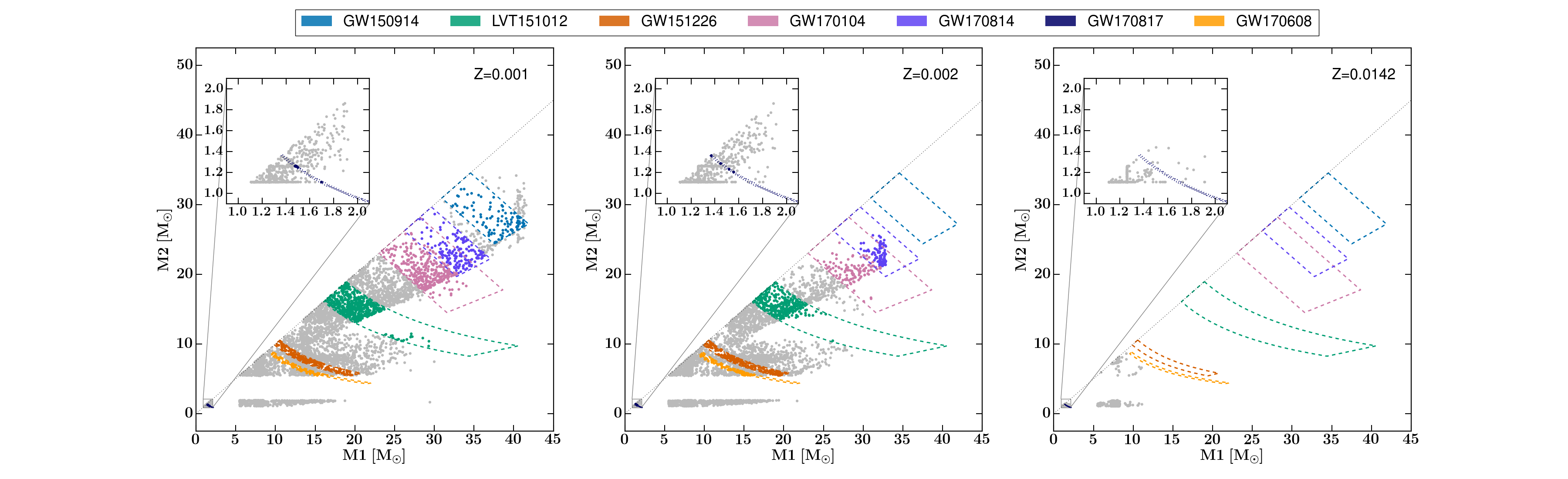}
	\caption{Masses of merging compact binaries predicted by the \texttt{Fiducial} model at three different metallicities: $\rm Z=0.001$ (left), $\rm Z=0.002$ (center) [cf.~\citep{stevenson2017formation}] and solar metallicity $\rm Z=0.0142$ (right). Coloured regions correspond to masses matching advanced LIGO detections within the reported 90 per cent credible intervals.}
	\label{fig:GWs}
\end{figure*}

\subsubsection{On the M\"uller prescription}
\label{sec:Muller}
We introduce the ``M\"uller" (04) explosion prescription as fitting formulae to the detailed models described by \cite{mueller2016simple}. The full description and fitting formulae are provided in Appendix \ref{sec:AppMuller} for use in other population synthesis studies. The ``M\"uller" prescription maps a CO core mass to a \ac{NS} remnant mass and a natal kick. The remnant and ejecta mass and the explosion energy are obtained semi-analytically and calibrated to numerical models. We update the analytic supernova models of \cite{mueller2016simple} by using a shock radius factor $\alpha_\mathrm{turb}=1.18$ and a compression ratio at the shock $\beta=3.2$, which fit constraints on the progenitor masses of Type~IIP supernovae \citep{smartt2015} slightly better than the original version. The natal kick velocity is obtained from these by assuming a uniform ejecta anisotropy \citep{janka2017}.  The natal kick magnitude, with a dominant mode at $v_{\rm{kick}}\approx 100~\rm km~s^{-1}$ is therefore correlated with the \ac{NS} remnant mass, unlike for the other models considered here. 
The mass range of \acp{NS} in our evolved population, using the ``M\"uller" explosion mechanism, is $[m_{\rm{NS,min}}, m_{\rm{NS,max}}]=[\minMassMuller,\maxMassMuller]~\rm M_{\odot}$. The formation rate and log~Bayes~factor of model (04) are $\rate=\formationVarFour$ per Milky Way equivalent galaxy per Myr  and $\log~\bayesFactor=\bayesVarFour$. This Bayes factor was calculated using only the $P-e$ distribution. The mass distribution (Figures \ref{fig:bnskde} and \ref{fig:chirpMass}) will play an important role in distinguishing the ``rapid'' (01), ``delayed'' (03) and ``M\"uller'' (04) explosion mechanism variations.

\subsubsection{On the Comparison with $\rm COMPAS\_\alpha$}
Here we compare our \Fiducial~model to the one described by \citet[][(00), $\rm COMPAS\_\alpha$]{stevenson2017formation}. The latter uses different parameterisations: both \acp{CCSN} and \acp{USSN} natal kicks are drawn from a high mode Maxwellian distribution and all \acp{ECSN} have a $v_{\rm kick}=0~\rm km~s^{-1}$; stability is determined using $\zeta_{\textrm{SPH}}$ for all stellar phases, which often leads to dynamically unstable mass transfer, particularly during case~BB \ac{RLOF}; and the binding energy parameter is $\lambda_{\textrm{fixed}}=0.1$ for all stars in any evolutionary stage. 

That study was successful in explaining all gravitational--wave events from the first advanced LIGO observing run 
\citep[GW150914, LVT151012 and GW151226][]{GW150914, O1:BBH,GW151226} 
via a single evolutionary scenario: isolated binary evolution. However, the same assumptions fail to reproduce the observed Galactic \ac{DNS} populations (see left panel of Figure~\ref{fig:PeMain}). Model (00), which yields a \ac{DNS} formation rate of $\rate_{\rm 00}=\formationVarZero$ per Milky Way equivalent galaxy per Myr, is the least preferred model from our variations, with a log~Bayes~factor of $\log~\bayesFactor=\bayesVarZero$. In particular, the extreme hardening of case~BB binaries through a second \ac{CE} phase in $\rm COMPAS\_\alpha$ leads to a gap in the $P-e$ distribution where systems such as J0737-3039 are observed. From the major changes, dynamical stability during case~BB mass transfer and a bimodal natal kick distribution are preferred over the alternatives in the \Fiducial~model (see unstable case~BB mass transfer (02) and single mode natal kick distribution (05) variations), which are ruled out in our model comparison.

On the other hand, the \Fiducial~model is able to explain, in a consistent form with \cite{stevenson2017formation}, the gravitational--wave events from the first advanced LIGO observing run, as well as GW170104 \citep{GW170104}, GW170608 \citep{GW170608}, GW170814 \citep{GW170814} and the \ac{DNS} merger GW170817 \citep{GW170817}, all detected during the second observing run of advanced LIGO and advanced Virgo (see Figure~\ref{fig:GWs}).

\subsubsection{On the Circularisation During Mass Transfer}
Our \Fiducial~model does not circularise the orbit during a mass transfer episode, except as a consequence of dynamically unstable mass transfer (\ac{CE}). As a variation, we consider circularisation at the onset of \ac{RLOF} (e.g. as a consequence of tidal dissipation prior to mass transfer or during the episode). We allow for two types of circularisation: 
\begin{enumerate*}[label=(\roman*), itemjoin={{, }}, itemjoin*={{, and }}]
  \item circularisation to periastron $a_{p}=a(1-e)$, which dissipates both orbital energy and angular momentum (12)
  \item circularisation to semilatus rectum $a_{SR}~=a(1-e^2)$, which conserves the angular momentum of the orbit (13).
  \end{enumerate*}
The \ac{DNS} formation rates and log~Bayes~factors are $\{\rate_{\textrm{12}},\rate_{\textrm{13}}\}=\{\formationVarTwelve,\formationVarThirteen\}$ per Milky Way equivalent galaxy per Myr and $\log~\{\bayesFactor_{\textrm{12}},\bayesFactor_{\textrm{13}}\}=\{\bayesVarTwelve,\bayesVarThirteen\}$ respectively. Rates decrease by less than a factor of $2$. Circularisation to periastron at the onset of mass transfer is slightly preferred than the alternatives, but not enough for us to consider it clearly preferred over the \Fiducial~model.  Circularisation which conserves angular momentum is not favoured or disfavoured with respect to the \Fiducial~assumption (i.e. no circularisation at all).

\subsubsection{On the Angular-Momentum Loss During Non-Conservative Mass Transfer}
During a non-conservative mass transfer episode, the specific angular momentum of the removed matter is determined by how mass leaves the system.
In our \Fiducial~assumption, any non-accreted mass is removed isotropically in the reference frame of the accretor; this mass loss mode is usually referred to as ``isotropic re-emission'' (01). Another common parameterisation is the ``Jeans" mode (14), which consists of ejecting the mass isotropically in the reference frame of the donor, similarly to fast winds. The last possibility we take into account is the formation of a circumbinary disk (15), with a radius of $a_{\rm disk}= 2a$ \citep{artymowicz1994dynamics}, from which the mass will be ejected. While isotropic re-emission (01) and the ``Jeans" mode (14) tend to effectively widen the orbit, that is not the case if mass is lost from a circumbinary disk (15). The formation rates of Galactic-like \acp{DNS} and the log~Bayes~factor are $\{\rate_{\textrm{14}},\rate_{\textrm{15}}\}=\{\formationVarFourteen,\formationVarFifthteen\}$ per Milky Way equivalent galaxy per Myr and $\log~\{\bayesFactor_{\textrm{14}},\bayesFactor_{\textrm{15}}\}=\{\bayesVarFourteen,\bayesVarFifthteen\}$ respectively. 
The \Fiducial~model is strongly preferred over the ``Jeans" mode (14) variation; it is also mildy preferred over the circumbinary disk (15) variation.
The mass loss mode also affects the future fate of the formed \acp{DNS}. The fraction of all formed \acp{DNS} that will merge in a Hubble time is $\{f_{01},f_{14},f_{15}\}=\{\fractionVarOne,\fractionVarFourteen,\fractionVarFifthteen\}$ for the ``isotropic re--emission'', ``Jeans'' and ``circumbinary disk'' mode, respectively.

\subsubsection{On the Common Envelope Parameters}
We consider several variations to the parameters that govern \ac{CE} evolution: $\lambda$, which determines the envelope binding energy, and $\alpha$, which determines the amount of orbital energy needed to expel the envelope. In our \Fiducial~model all of the \acp{DNS} experience a \ac{CE} phase and therefore varying $\lambda$ and $\alpha$ from the \Fiducial~model choices (i.e. $\lambda_{\textrm{Nanjing}}$ and $\alpha=1$) will affect the final distributions.  

$\lambda_{\textrm{Nanjing}}$ is a function of core mass, total mass and radius. We use a fixed value $\lambda_{\textrm{fixed}}=0.1$ (08) for comparison with previous population synthesis studies  \cite[e.g.,][]{belczynski2002comprehensive}. Recently, \cite{Kruckow:2016tti} found for several models at different mass and metallicity that $\lambda$ depends on the radius in a roughly power-law form $\lambda \propto R^{\beta}$, with $-1\leq\beta\leq-2/3$. We made a fit to Figure~1 of \cite{Kruckow:2016tti} in the form $\lambda_{\textrm{Kruckow}} = 1600\times0.00125^{-\beta}R^{\beta}$, assuming a monotonically decreasing function. For our particular variation, we use an average value where $\beta=-5/6$ (09). The formation rates of \acp{DNS} and the log~Bayes~factors for these variations in $\lambda$ are $\{\rate_{\textrm{08}},\rate_{\textrm{09}}\}=\{\formationVarEight,\formationVarNine\}$ per Milky Way equivalent galaxy per Myr and $\log~\{\bayesFactor_{\textrm{08}},\bayesFactor_{\textrm{09}}\}=\{\bayesVarEight,\bayesVarNine\}$ respectively, not favouring nor disfavouring the $\lambda$ variations with respect to the \Fiducial~model.

Higher values of $\alpha$ lead to wider post--\ac{CE} orbits than low values of $\alpha$.  Without exploring the full and continuous parameter space, we vary $\alpha$ to extreme values of $\alpha_{min}=0.1$ (10) and $\alpha_{max}=10$ (11). Values of $\alpha>1$ suppose that there are substantial additional energy sources, such as recombination energy and/or nuclear energy \citep{podsiadlowski2010explosive, ivanova2013common}  that contribute to the energy budget for \ac{CE} ejection, in addition to the orbital energy. The extreme value of $\alpha_{max}=10$ is more for illustration purposes rather than to mimic a particular physical interaction; in this case $\alpha_{max}=10$ can only be explained if it comes from nuclear energy. The formation rates of \acp{DNS} and the log~Bayes~factors for variations in $\alpha$ are $\{\rate_{\textrm{10}},\rate_{\textrm{11}}\}=\{\formationVarTen,\formationVarEleven\}$ per Milky Way equivalent galaxy per Myr and $\log~\{\bayesFactor_{\textrm{10}},\bayesFactor_{\textrm{11}}\}=\{\bayesVarTen,\bayesVarEleven\}$ respectively, not clearly favouring nor disfavouring the $\alpha$ variations with respect to the \Fiducial~model. 
The choice of $\alpha$ influences not only the number of created \acp{DNS}, but also the number of mergers. The fraction of all formed \acp{DNS} that will merge in a Hubble time is $\{f_{01},f_{10},f_{11}\}=\{\fractionVarOne,\fractionVarTen,\fractionVarEleven\}$. 

Additionally, we also consider the ``pessimistic" \ac{CE} assumption (19). This assumption yields a \ac{DNS} population which is a subset of the population under the \Fiducial~model, with binaries that enter the \ac{CE} while the donor is classified as a \ac{HG} star removed, as these are assumed to always lead to merger.  The ``pessimistic'' \ac{CE} assumption (19) is therefore expected to decrease \ac{DNS} formation rates. The formation rates of \acp{DNS} and the log~Bayes~factors for these variations are $\{\rate_{\textrm{01}},\rate_{\textrm{19}}\}=\{\formationVarOne,\formationVarNineteen\}$ per Milky Way equivalent galaxy per Myr and $\log~\{\bayesFactor_{\textrm{01}},\bayesFactor_{\textrm{19}}\}=\{\bayesVarOne,\bayesVarNineteen\}$ respectively. The likelihood of the ``pessimistic'' model (19) is similar to the one from the \Fiducial~model, which means the $P-e$ distribution alone is insufficient to pick between these models.  Additional constraints, such as merger rates, would be needed to determine the preferred model.

\subsubsection{On the Effect of Thermal Eccentricity}
The only initial distribution we varied in this study was eccentricity. In order to simulate a population with non circular binaries at ZAMS we use the thermal eccentricity distribution (16), which has the form of $f_{\textrm{e}}(e)= 2e$ \citep{heggie1975binary}. In this variation, the first episode of mass transfer begins once the primary expands to fill its Roche lobe at periastron.  This changes the range of initial periods leading to interaction \citep{deMinkBelczysnki2015}.

The formation rate and log~Bayes~factor of model (16) are $\rate=\formationVarSixteen$ per Milky Way equivalent galaxy per Myr  and $\log~\bayesFactor=\bayesVarSixteen$ respectively. While formation rates drop by a factor of approximatively 3, the $P-e$ distribution of forming \acp{DNS} is not significantly affected. The drop in the formation rate is due to enhanced rates of interactions of \ac{MS} stars that only need to fill their Roche lobe at periastron; if that mass transfer episode is unstable, the two \ac{MS} stars merge.

\subsection{On Mass Ratio Distributions}
\label{subsec:massRatio}
Figure~\ref{fig:qCDFs} shows the impact of the choice of the \ac{SN} remnant mass model on the \ac{DNS} mass ratio distributions. 
The \Fiducial~model shows two distinct peaks in the mass ratio distribution around $\rm q_{\rm DCO}=0.87$ and $\rm q_{\rm DCO}=1$. The two peaks can be explained given the evolution of \textit{Channel~I} and \textit{Channel~II}, respectively. For the full discussion on the characteristics of the mass ratio for the \Fiducial~model, see Section \ref{subsubsec:FiducialMassRatio}.

In the ``delayed'' prescription (03) most of the \acp{USSN} change mass from $\minMassRapid~\rm M_{\odot}$ to $1.28~\rm M_{\odot}$, with respect to the ``rapid'' mechanism; therefore, the mass ratio of systems where the primary collapsed in an \ac{ECSN} and the secondary in an \ac{USSN} approaches 1, yielding an even more dominant peak at $q_{\rm DCO}=1$ in the overall mass ratio distribution. \textit{Channel~II} leads to the second peak, with mass ratio $q_{\rm DCO}=1$, as in the \Fiducial~model. This results in a cumulative distribution function for the ``delayed" mechanism (03) with a mass ratio between $\minMassRatioDelayed \leq q_{\rm DCO} \leq 1$, where \qAboveEightyDelayed \  per cent of the systems have $q_{\rm DCO} > 0.80$, \qAboveNinetyDelayed\ per cent have $q_{\rm DCO} > 0.90$ and \qAboveNinetyFiveDelayed \  per cent have $q_{\rm DCO} > 0.95$.

The remnant masses in the M\"uller prescription (04), as shown in Figure~\ref{fig:bnskde} and \ref{fig:Muller}, have a wider spread and vary more at the low mass end. In this model, there is no significant pile-up. There is more scatter, with \qAboveEightyMuller \  per cent of the systems having $q_{\rm DCO} > 0.8$, \qAboveNinetyMuller \  per cent having $q_{\rm DCO} > 0.9$ and \qAboveNinetyFiveMuller \  per cent having $q_{\rm DCO} > 0.95$. 

\subsection{On the Chirp Mass Distribution}
\label{subsec:chirpMass}
\begin{figure}
	\centering
		\includegraphics[trim={2.2cm 7cm 3.7cm 7.4cm},clip,scale=0.55]{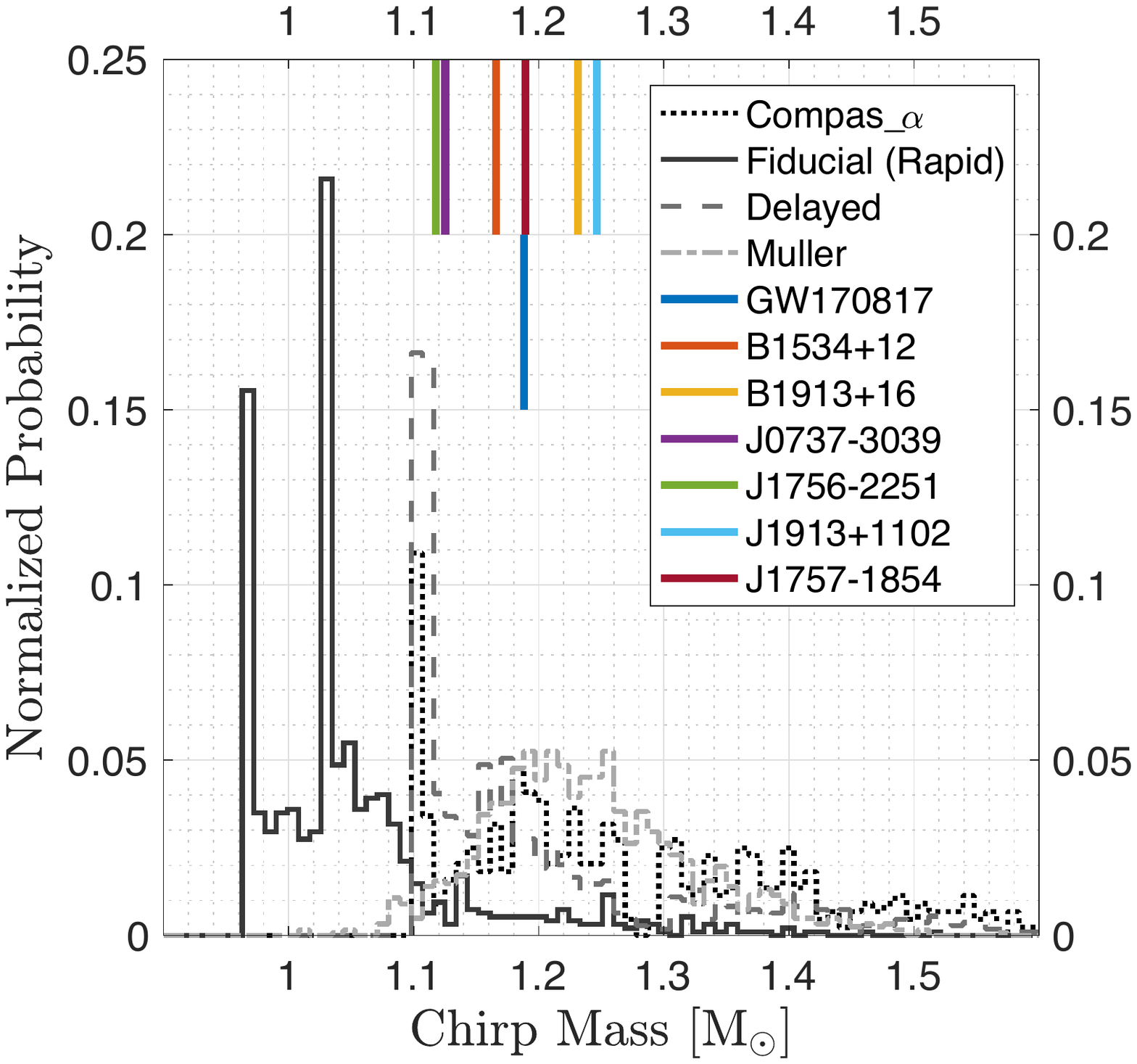}
	\caption[]{
	Chirp mass distribution of \acp{DNS} with a delay time smaller than a Hubble time: (00) $\rm COMPAS\_\alpha$ [black dotted], (01) \Fiducial~Fryer Rapid [dark gray solid], (03) Fryer Delayed [gray dashed] and (04) M\"uller [light gray dot-dashed]. Galactic \acp{DNS} with an estimated delay time smaller than a Hubble time are indicated at the top. GW170817, the only GW signal detected from \acp{DNS} to date, is shown as a vertically offset thick green line, with a similar chirp mass ($1.188~\rm M_{\odot}$) as J1757-1854 in purple. All systems have precise mass measurements with error bars within the thickness of the line. 
	} 
	\label{fig:chirpMass}
\end{figure}

Figure~\ref{fig:chirpMass} shows the chirp mass distributions of \acp{DNS} which will merge within a Hubble time. We compare the prediction of our \Fiducial~model (01) which uses the ``rapid'' explosion mechanism, to the model which uses the ``delayed'' (03) explosion mechanism and to that which uses the ``M\"uller'' (04) prescription.

Additionally, we also show the $\rm COMPAS\_\alpha$ (00) chirp mass distribution which uses the ``delayed'' mechanism. As expected, the chirp mass distributions show similarities with the mass ratio distributions, reproducing the same sharp features (peaks) explained in Section \ref{subsec:massRatio}. In Figure~\ref{fig:chirpMass} we added all the confirmed \acp{DNS} with an estimated delay time smaller than the Hubble time, as well as GW170817. 

We find that the ``rapid'' (01) mechanism predicts that most of the \acp{DNS} will have chirp mass lower than J1756-2251, which has the lowest chirp mass among confirmed  \acp{DNS} with good mass constraints. In fact, the ``rapid'' \ac{SN} mechanism (01) allows for low-mass \acp{NS} which would be difficult to differentiate from \ac{NS}--white dwarf binaries; there are several non-confirmed \acp{DNS} or poorly constrained \ac{DNS} masses in the region favoured by the ``rapid'' mechanism (01) \citep{Ozel:2010,ozel2016masses}. On the other hand, the seven existing well-constrained mass measurements in this study are inconsistent with the predictions of the \Fiducial~model (01) at a $> 4\sigma$ level.  None of these seven measurements fall below a chirp mass of 1.1 $\rm M_{\odot}$, while $83$ per cent of \acp{DNS} in the \Fiducial~model have lower chirp masses. This suggests that the ``rapid'' mechanism under-predicts the amount of collapsed mass for the lowest-mass \acp{NS} for both \acp{ECSN} and \acp{USSN}. 

All other \ac{SN} prescriptions considered here yield \ac{DNS} chirp mass distributions starting above 1.1 $\rm M_{\odot}$. Unsurprisingly, the ``delayed'' mechanism (03) has a very similar distribution to $\rm COMPAS\_\alpha$ which uses the same explosion mechanism. They both predict systems matching all chirp masses (see Figure 10), with a peak close to the lowest observed \ac{DNS} chirp masses, J1756-2251 and J0737-3039.  The ``M\"uller'' prescription (04) yields a similarly broad chirp mass distribution above 1.1 $\rm M_{\odot}$. The ``delayed'' (03) and ``M\"uller'' (04) \ac{SN} fallback prescriptions cannot be distinguished based on existing mass measurements.  However, the separation of $\approx 0.4$ between the predicted chirp mass cumulative distribution functions for these two models suggests that $\sim$ 10 additional chirp mass measurements (whether from radio pulsars or merging \acp{DNS}) would be sufficient to tell these models apart.

\subsection{On Kicks}
When binaries survive a \ac{SN} explosion, they may get significant centre-of-mass kicks from both natal \ac{NS} kicks and Blaauw recoil \citep{blaauw1961origin} from mass loss. The resulting \ac{DNS} population should therefore be more broadly spatially distributed in the Galaxy than the regions of massive star formation. We follow a population of \Fiducial~model \acp{DNS} with the predicted total kick distribution in a Galactic potential starting from birth in the thin disk.  While we find that, as expected, kicks broaden the distribution of Galacto-centric distances (see Figure~\ref{fig:kicks} in appendix \ref{sec:AppD}, where the details of this analysis are presented), the deep Galactic potential well means that this broadening is relatively small and challenging to test for. In practice, the spreading of \acp{DNS} away from the thin disk may be even smaller than estimated here, because our simplified case~BB mass loss models imply fairly high remaining core masses, between $1.6 \leq m_{CO}/\rm{M_{\odot}} \leq 4.6$, while detailed calculations of ultra-stripping suggest lower remnant core masses $1.45 \leq m/\rm{M_{\odot}} \leq 3.15$ \citep{tauris2015ultra}. Reducing COMPAS core masses in line with \citet{tauris2015ultra} would both reduce Blaauw kicks and \ac{DNS} eccentricities.  On the other hand, three quarters of short \acp{GRB} are found outside the effective radius of the host galaxy \citep{FongBerger:2013}, providing a strong constraint on the binary natal kick distribution; \citet{FongBerger:2013} estimate total kicks of $\approx 20$ -- $140$ km s$^{-1}$.

\subsection{On Rates}
\label{subsec:Rates}
\subsubsection{DNS Merger Rates}
\ac{DNS} formation and merger rates are constrained by the observed sample of Galactic binary pulsars \citep[e.g.,][]{Kim:2003kkl, OShaughnessyKim:2009}, by observations of short \acp{GRB} \citep{FongBerger:2013}, and will ultimately be measured with gravitational--wave observations \citep[see][for a review]{MandelOShaughnessy:2010}. Rates inferred from Galactic binary pulsars are dominated by a few systems and are sensitive to the imperfectly known pulsar radio luminosity distribution \citep{Kalogera:2004tn}. Short \acp{GRB} extend the observations beyond the Milky Way to cosmological distances, but inference from these is complicated by the difficulty of measuring jet opening angles and uncertain selection effects, and relies on the additional assumption of a one-to-one correspondence between short \acp{GRB} and \ac{DNS} mergers \citep{2014ARA&A..52...43B}.
\cite{ratesdoc} combined the existing observational constraints to suggest that the \ac{DNS} merger rate lies between $1$ and $1000$ $\rm Myr^{-1}$ in a Milky Way equivalent galaxy (approximately $10$ to $10000$ Gpc$^{-3}$ yr$^{-1}$), with a likely value toward the middle of this range. All of the models presented here fall within this range, although we focus on the Milky Way \ac{DNS} population rather than the merger rate, and hence did not consider the convolution of the \ac{DNS} formation rate and delay time distribution over cosmic history. 

Other recent population synthesis studies give estimates that, like ours, fall in the two lower decades of this range. \citet{chruslinska2018double} use \texttt{StarTrack} to predict a local merger rate density of $48$ Gpc$^{-3}$ yr$^{-1}$ for their standard assumptions and $600_{-300}^{+600}$ Gpc$^{-3}$ yr$^{-1}$ for a very optimistic set of assumptions. \citet{belczynski2017GW170817} also use \texttt{StarTrack} to argue that even these rates are 2 orders of magnitude larger than the contribution from globular or nuclear clusters.	 \citet{2018arXiv180105433K} use $\rm C_{OM}B_{IN}E$ to predict an upper limit of local merger rate of $400$ Gpc$^{-3}$ yr$^{-1}$.

Meanwhile, \citet{GW170817} estimate a \ac{DNS} merger rate of $1540^{+3200}_{-1220}$ Gpc$^{-3}$ yr$^{-1}$ based on GW170817 alone.  However, given the significant Poisson uncertainty and sensitivity to rate priors from a single observation\footnote{For example, shifting from a flat-in-rate prior to a $p(R) \propto 1/\sqrt{R}$ Jeffreys prior \citep{Jeffreys:1946}  would reduce the peak of the posterior by a factor of 2 following one detection.  Furthermore, the posterior peak is a factor of $1.67$ lower than the posterior median quoted by \cite{GW170817}.}, the addition of this one (albeit, very special) event to the population of merging Galactic \acp{DNS} and short \acp{GRB} does not significantly shift the observational constraints on the \ac{DNS} merger rate.  In fact, given the similarity of the predicted \ac{DNS} formation rates among most models presented here, observational constraints on the rate alone will not be sufficient to distinguish between these models in the near future.   

\subsubsection{Supernova Rates}
We estimate the \ac{SN} rates for our \Fiducial~model (01). Given the ambiguity in \ac{SN} classification, we make simplifying assumptions to convert our models into observational predictions.  We consider all progenitors with a hydrogen envelope to lead to hydrogen rich \acp{SN} (type II excluding type IIb) and the rest are considered stripped \acp{SN} (either hydrogen absent type Ib or Ic or hydrogen poor type IIb). Our total rate of \acp{SN} leading to \ac{NS} formation  is \CCSNperMsolSF~per~$\rm M_{\odot}$, which includes both \acp{ECSN} and \acp{USSN}.  Among these, \typeIItoCCSN~per cent are hydrogen rich and the remaining \typeItoCCSN~per cent are classified as stripped \acp{SN}, including all \acp{USSN}.  We predict that \acp{USSN} that follow after case~BB mass transfer onto a \ac{NS} companion should make up \USSNtoTypeI~per cent of all stripped \acp{SN} and \USSNtoCCSN~per cent of all \acp{SN} leading to \ac{NS} formation.  

Our total \ac{SN} rate prediction is consistent with \cite{zapartas2017ccsne}, a population synthesis study which reports \ac{CCSN} rates in binaries between $0.0035$--$0.0253$ per~$\rm M_{\odot}$, depending on the assumed IMF. Our estimates for the fraction of stripped \acp{SN} compare well with observational results. \citet{2013MNRAS.436..774E} find that the fractions of hydrogen rich and stripped \acp{SN} leading to \ac{NS} formation are $61.9$ and $38.1$ per cent respectively; that study was made with \acp{SN} discovered between $1998$ and $2012$ in galaxies with recessional velocities less than $2000~\rm km~s^{-1}$. More recently, \cite{2017PASP..129e4201S} report that $69.6$ per cent of \acp{CCSN} are hydrogen rich (according to the definition above), while the remaining $30.4$ per cent come from stars with stripped envelopes. 

\section{Discussion \& Conclusions}
\label{sec:discussion}
We used the COMPAS rapid population synthesis code to follow the evolution of massive stellar binaries and thus generate a population of \acp{DCO}.  We quantitatively validated our models by comparing the predicted $P-e$ distribution of \acp{DNS} against the observed Galactic \ac{DNS} distribution, and qualitatively compared the predicted rate and mass distribution of Galactic \acp{DNS} to observations.  We considered variations relative to the \Fiducial~model in order to investigate the impact of uncertain evolutionary physics. We find that:

\begin{itemize}

\item{\it Case BB mass transfer during \ac{DNS} formation must be predominantly stable.}  We considered the possibility that \ac{HeHG} of the secondary leads to dynamically unstable mass transfer and a second \ac{CE} phase \citep{dewi2003late} in Variation (02). In fact, this was our initial default model, consistent with $\rm COMPAS\_\alpha$ (00) in this assumption.  However, the lack of \acp{DNS} with few-hour orbital periods (such as J0737-3039) in this variation (see Figure~\ref{fig:PeMain}), as well as our Bayesian analysis, indicates that most case~BB mass transfer episodes must be stable.  This finding is consistent with the detailed models of \cite{tauris2015ultra}.  However, some case~BB dynamically unstable systems could exist without being detectable in the observed \ac{DNS} population: the very short orbital periods of \acp{DNS} that were hardened by two \ac{CE} phases would lead them to merge in less than a few hundred thousand years. While our study assumes constant star formation within the history of the Galaxy, the short orbital period \acp{DNS} would be disfavoured in Galactic star formation history models without recent periods of starbursts.

\item{\it A bimodal \ac{SN} natal kick distribution is preferred over a single mode one.}  We find that a bimodal natal kick distribution (with non-zero components) with lower natal kicks for \acp{ECSN} and \acp{USSN} and higher natal kicks for standard \acp{CCSN} is preferred (see variations (05), (06), (07)). If \acp{ECSN} and/or \acp{USSN} are given the high natal kicks consistent with the observed velocities of isolated pulsars \citep{hansen1997pulsar,hobbs2005statistical}, wider binaries are overwhelmingly disrupted by \acp{SN}, and observed wide \acp{DNS} cannot be reproduced in the models. A bimodal \ac{SN} natal kick distribution is consistent with the findings of other population synthesis studies (see \cite{pfahl2002new} and \cite{belczynski2002comprehensive} as well as with comparison to observations from \cite{schwab2010further}, \cite{beniamini2016formation} and \cite{Verbunt2017bimodal}); although \citet{Oshaughnessy:2008} didn't find evidence for multiple natal kick distributions. 
\\
	
The aforementioned findings in our paper, stability during case~BB mass transfer and a bimodal natal kick distribution, are broadly in agreement with those in \citet{andrews2015evolutionary}, which used a smaller sample of eight Galactic \acp{DNS} instead of the current 14 confirmed systems and carried out population synthesis by mainly varying \ac{CE} parameters and natal kick magnitudes. \citet{andrews2015evolutionary} find that it is likely that short-period low-eccentricity systems went through an evolutionary channel which includes stable case~BB mass transfer. Their study also points out that the cores of \ac{ECSN} progenitors should have relatively low mass, which can be related to lower natal natal kick magnitude.
\\

\item{\it Predicted \ac{DNS} formation rates across variations are consistent with observations.} The formation rate of \acp{DNS} in the \Fiducial~model is 24 Myr$^{-1}$ in the Milky Way. The Milky Way \ac{DNS} formation rate for all considered variations is 5 -- 31 Myr$^{-1}$. All rates are consistent with observations \citep{ratesdoc}, including the inferred rate from the GW170817 gravitational--wave detection \citep{GW170817}, and cannot be used to differentiate between the models at this point.
\end{itemize}

We also considered multiple \ac{SN} explosion mechanisms, including varying the fallback mass (Fryer ``rapid'' (01) and Fryer ``delayed'' (03) variations) and a coupled mass--kick model calibrated to numerical simulations (``M\"uller'' (04) prescription).

Low-mass iron-core \acp{CCSN} may have reduced natal kicks, but are given standard \ac{CCSN} natal kicks in the Fryer models, including the \Fiducial~model. The mass distribution of observed systems is not consistent with the very low masses predicted by the Fryer ``rapid'' fallback prescription  used in the \Fiducial~model (01). Furthermore, observations do not show a peak in the mass distribution around $1.26~\rm M_\odot$ where \acp{ECSN} should fall in our models. The remnant mass of an \ac{ECSN} depends on the \ac{NS}'s equation-of-state and indicates either that \acp{ECSN} are less common in binaries than we expected or that the \ac{ECSN} models should be revisited, as similarly noticed by \cite{2018arXiv180105433K}. With only $\sim 10$ additional \ac{DNS} mass measurements it will be possible to further constrain the \ac{SN} fallback mechanisms, distinguishing between the ``M\"uller'' (04) and Fryer ``delayed'' (03) variants, both of which are consistent with existing observations.


Further input on natal kick velocity distributions should come from a better comparison with observed isolated pulsar natal kicks.  At the moment, the observed isolated pulsar distribution is used to calibrate the \ac{CCSN} natal kicks in binaries.  However, the sample of observed isolated pulsars is contaminated by pulsars from disrupted binaries. Therefore, the approach we used here, which is also used by most population-synthesis codes, is not self-consistent: the observed single-pulsar velocity distribution should be checked for consistency against a model which includes contributions from both single and binary massive stars.  In particular, observations should be tested for evidence of the predicted low natal kicks associated with \acp{ECSN}, which may preferentially occur in binaries \citep{podsiadlowski2004effects} that may subsequently be disrupted.

We assumed a solar metallicity $\rm Z_{\odot}=0.0142$ for massive stars in the Galaxy.  In reality, the Galaxy has a distribution of metallicities at the present day, as well as a history of metallicity evolution over time, since present-day \ac{DNS} systems and particularly \ac{DNS} mergers may have formed at earlier times or in lower-metallicity regions \citep[see][for a discussion of Galactic binary black hole formation]{lamberts2018predicting}.  While Figure~\ref{fig:GWs} confirms that, for a suitable choice of metallicity and initial conditions, the \Fiducial~model can produce compact binary mergers with masses matching all of the existing gravitational--wave observations; it also demonstrates that metallicity does impact the rate and properties of merging \acp{DNS}.  Therefore, the metallicity-specific star formation history of the Milky Way could affect the details of the modelled \ac{DNS} population.

We do not account for selection effects in the observed Galactic \ac{DNS} population in this study; see \cite{tauris2017formation} for a detailed discussion. Binaries with very short orbital periods may be selected against because of the orbital acceleration of the pulsar, which changes the apparent spin period; they will also have short merger times, and their location within the Galaxy will be sensitive to the details of recent star formation history.  Meanwhile, binaries with extremely long orbital periods may also be challenging to detect, since they are less likely to be recycled during binary evolution, and detectable radio emission from non-recycled pulsars is expected to last for $\lesssim 50$ Myrs \citep{2004hpa..book.....L}.


The \ac{DNS} formation models presented here can also be tested against observable populations of massive stars during intermediate phases before \ac{DNS} formation.  Neutron star Be/X-ray binaries \citep[e.g.,][]{knigge2011two} offer a particularly promising test case; for example, the observed correlation between the orbital period and the \ac{NS} spin, with the latter appearing to be bimodal, could indicate distinct \ac{SN} classes in their evolutionary history \citep{knigge2011two}. Spin distribution predictions could also be compared to observed pulsar spin periods in both isolated pulsars \citep[e.g.][]{Kiel:2008xw} and in \ac{DNS} systems \citep[e.g.][]{2005MNRAS.363L..71D,2011MNRAS.413..461O,tauris2017formation}. However, determining the \ac{NS} spin-up or spin-down through binary interactions and pulsar evolution requires additional modelling assumptions, and hence spin models were not included in the present study. Meanwhile, more detailed studies of natal kicks in the Galactic potential could lead to additional constraints on natal kick distributions.  Moreover, gravitational--wave detections will produce an ever larger catalogue of accurate mass measurements, at least for the chirp mass parameter. Together, these growing observational data sets will enable increasingly accurate tests of the massive stellar binary evolution models described here. 


\section*{Acknowledgments}
AVG acknowledges support from CONACYT and thanks the PHAROS COST Action (CA16214) for partial support.
CN and IM acknowledge partial support from the STFC.
KB acknowledges support from the Polish National Science Center
(NCN) grants Sonata Bis 2 (DEC-2012/07/E/ST9/01360) and OPUS (2015/19/B/ST9/01099).  
SJ is partially supported by the Strategic Priority Research Program of the Chinese Academy of Sciences ``Multi-waveband Gravitational Wave Universe'' (Grant No.~XDB23040000), and is also grateful to the Chinese Academy of Sciences (President's 
International Fellowship Initiative grant no. 2011Y2JB07), and the National Natural
Science Foundation of China (grant no. 11633005).
SdM acknowledges the European Union's Horizon 2020 research and innovation programme
for funding from the European Research Council (ERC), Grant agreement No. 715063.
SdM and IM acknowledge the hospitality of the Kavli Institute for Theoretical physics, Santa
Barbara, CA. Their stay was supported by the National Science Foundation under Grant
No.\ NSF PHY11-25915.
BM was supported by the Australian Research Council through ARC Future Fellowship FT160100035. SS was supported by the Australian Research Council Centre of Excellence for Gravitational Wave Discovery (OzGrav), CE170100004.

We thank the Niels Bohr Institute for its hospitality while part of this work was completed, and acknowledge the Kavli Foundation and the DNRF for supporting the 2017 Kavli Summer Program. 
We also thank Jeff Andrews, Christopher Berry, Ross Church, David Stops, Jason Hessels, Serena Vinciguerra and Manos Zapartas for discussions, suggestions and assistance during the writing of this manuscript.

\bibliographystyle{mnras}
\bibliography{dns_bib}
		
\clearpage
\appendix
\section{Likelihood calculation}
\label{sec:likelihood}
\begin{figure*}
	\centering
	\includegraphics[trim={0.5cm 1.5cm 0.5cm 1.5cm},clip,width=\textwidth]{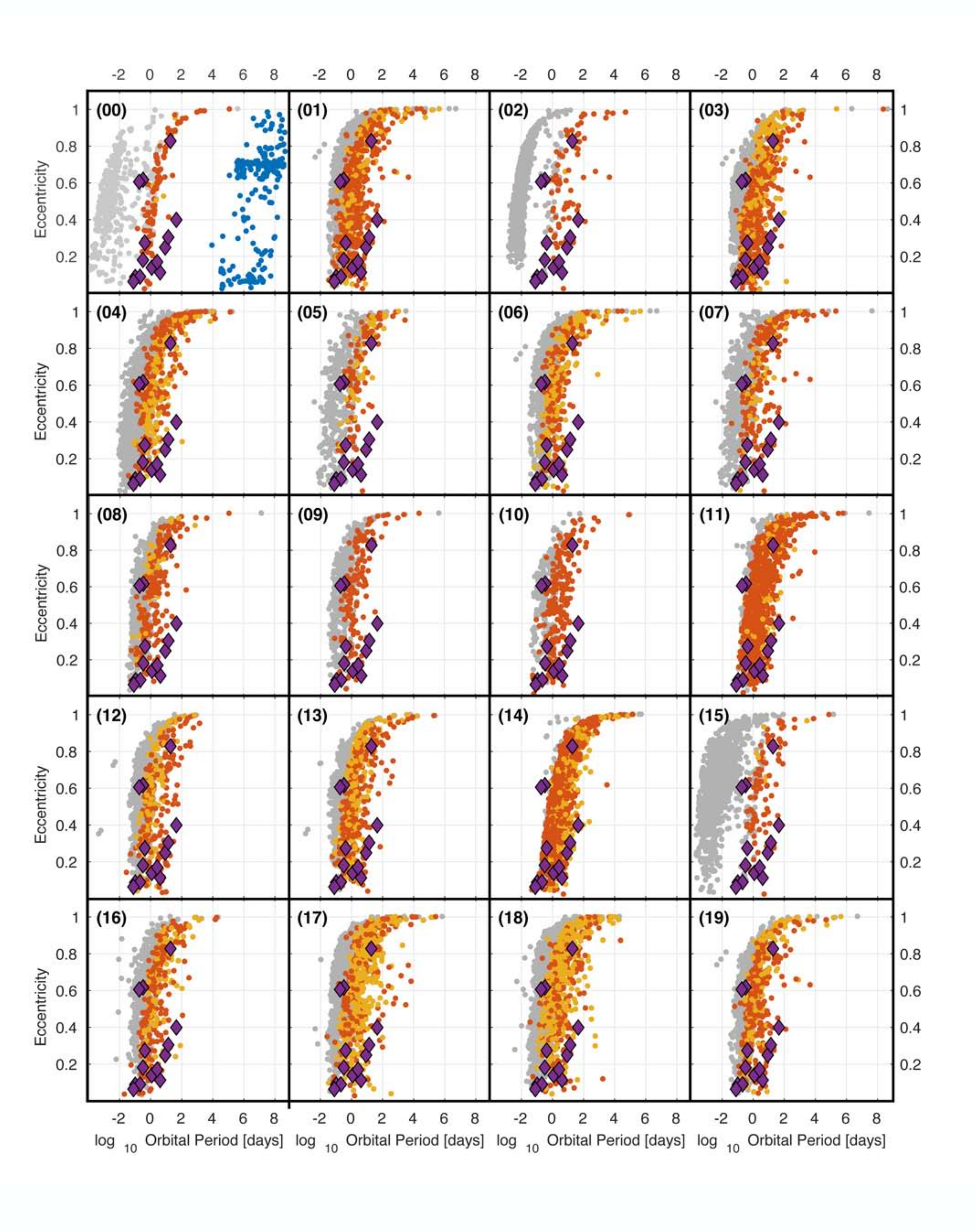}
	\caption{
Predicted $P-e$ distribution of Galactic \acp{DNS} under the \Fiducial~model. \textcolor{gray}{\textbf{Gray}} dots are all \ac{DNS} at \ac{DCO} formation. \ac{DCO} period and eccentricity are evolved forward from birth until present age given gravitational--wave radiation emission, likely removing short-lived short-period binaries from the observable population. Coloured dots represent the \ac{DNS} distribution at present age. Colour denotes the type of \ac{CE} phase: \lone{blue} for no \ac{CE} phase, \ltwo{red} for a single-core and \lthree{yellow} for a double-core \ac{CE} phase. The single-core and double-core can be, in most cases, associated with \textit{Channel~I} and \text{Channel~II} respectively (see Section \ref{subsubsec:FormationChannels}). \lfour{Purple} diamonds represent the observed Galactic \acp{DNS}; all observed systems have precise $P-e$ measurements with error bars within the thickness of the symbol. 
	} 
	\label{fig:nineteenpanels}
\end{figure*}

Our methodology follows \citet{andrews2015evolutionary}. We can write the base $e$ log-likelihood $\log \, \mathcal{L}$ as
\begin{equation}
\log \, \mathcal{L} = \sum_{b=1}^{N_\mathrm{obs}} \log \, p(\log P_b, e_b | M) ,
\label{eq:log_likelihood}
\end{equation}
where $e_{b}$ and $\log \, P_{b}$ are the eccentricity and log of the orbital period in days for the $b$-th observed \ac{DNS}, respectively; $N_\mathrm{obs}=14$ observations were used here (see Table I and associated discussion). The term $\, p( \log P_b, e_b | M) $ describes the likelihood of observing the $b$-th \ac{DNS} given a model $M$, where our models are described in Table~\ref{tab:models} and shown in Figure~\ref{fig:nineteenpanels}. We therefore need a way of calculating the 2D probability density given the discrete simulated \ac{DNS} binaries we have for each model.

We evolve the eccentricity and period of each simulated \ac{DNS} as it emits gravitational radiation according to \citet{peters1964gravitational}. We stop the inspiral evolution when the system either merges or reaches 10 Gyr (a proxy for the age of the Galactic thin disk, see \cite{2005A&A...440.1153D}). 
We place systems into linearly spaced bins in eccentricity, with the lowest bin spanning $e \in [0,10^{-4}]$, and determine the log period $\log P_k$ when the system enters each bin with eccentricity $e_k$ and the time the system spends in that bin $\Delta t_k$, which is subject to
\begin{equation}
\sum_{k} \Delta t_{k} = t_\mathrm{delay} .
\label{eq:sum_of_dt_equals_tdelay}
\end{equation}

We weigh the contribution of each binary at each point in its evolutionary history to the probability density map by $\Delta t_k$, since a system is more likely to be observed in the part of the orbit where it spends more of its time. Since tight, highly eccentric binaries evolve the fastest due to gravitational radiation, this has the effect of down weighting those binaries in our analysis (see Figure~\ref{fig:PeWG}). 

We construct the probability density map from a discrete sample of simulated binaries by means of a weighted kernel density estimator.\footnote{We found that density maps estimated via a 2D binned histogram, as used by \citet{andrews2015evolutionary}, were extremely sensitive to the chosen number of bins.}  We model the 2D probability density as a sum of weighted Gaussians
\begin{equation}
p(\log P, e | M) = \sum_{j=1}^{n_\mathrm{binaries}} \sum_{k=1}^{n_\mathrm{timesteps,j}} \frac{\Delta t_{k}}{T} N(\bm{\mu}_{k}, \bm{\Sigma}_{k}) ,
\label{eq:sum_of_weighted_gaussians}
\end{equation}
where 
\begin{equation}
T = \sum_{j=1}^{n_\mathrm{binaries}} \sum_{k=1}^{n_\mathrm{timesteps,j}} \Delta t_{k} = \sum_{j=1}^{n_\mathrm{binaries}} t_\mathrm{delay,j} ;
\end{equation}
$N(\bm{\mu}, \bm{\Sigma})$ is the 2D normal distribution with mean
\begin{equation}
\bm{\mu}_{k} = (\log \, P_{k}, e_k ) ,
\end{equation}
and the covariance $\bm{\Sigma}_{k}$ is chosen to be the same for all samples
\begin{equation}
\bm{\Sigma}_{k} = \begin{bmatrix}
    b_{\log P}^2 & 0 \\
    0 & b_{e}^2
\end{bmatrix} ,
\label{eq:covariance_matrix}
\end{equation}
where $b_{\log P}$ and $b_{e}$ are the `rule-of-thumb` \citep{silverman1986density} bandwidth parameters which determine how much we `smooth' the distribution.  We choose $e_\mathrm{max}=1$, $e_\mathrm{min}=0$, $\log \, (P_\mathrm{min}/\mathrm{days})=-6$ and $\log \, (P_\mathrm{max}/\mathrm{days})=4$ for our analysis.

The log-likelihoods fluctuate by $\mathcal{O}(1)$ depending on the choice of bandwidth.  This systematic uncertainty in the estimated likelihoods arises because our theoretical distributions are built from a finite number of samples, and could  be improved with larger simulation campaigns.


\clearpage
\section{Model for the Dependence of the Kick Velocity on Explosion Parameters}
\label{sec:AppMuller}
  The most viable mechanism for producing sizeable natal kicks in
  \ac{CCSN} explosions is the gravitational tug-boat
  mechanism, which relies on the acceleration of the \ac{NS} due
  to the net gravitational pull exerted by anisotropic ejecta during
  the first few seconds after shock revival
  \citep{scheck2004,scheck2006multidimensional,nordhaus2010,wongwathanarat2013}. 
\cite{bray2016} suggested that this natal kick could be correlated
with other explosion properties.
An attempt to clarify these correlations based on the phenomenology of multi-dimensional simulations was then made by \cite{janka2017}, whose natal kick estimate we briefly review here, since it largely agrees with the one we developed for COMPAS.
Invoking total momentum conservation, \cite{janka2017} considered the momentum $|\mathbf{p}_\mathrm{ej}|$ of the ejecta at a time when the natal kick asymptotes to its final value. Introducing an anisotropy parameter $\alpha_\mathrm{kick}$ to relate $|\mathbf{p}_\mathrm{ej}|$ to the spherical quasi-momentum of the ejecta as
\begin{equation}
\alpha_\mathrm{kick}=\frac{|\mathbf{p}_\mathrm{ej}|}{\int_\mathrm{ejecta} \rho |\mathbf{v}| \,dV},
\end{equation}
\cite{janka2017} then invoked dimensional analysis to relate the ejecta (and NS) momentum to the kinetic energy $E_\mathrm{kin}$ and mass  $m_\mathrm{ej}$ of the anisotropic ejecta behind the shock as
\begin{equation}
|\mathbf{p}_\mathrm{ej}| = \alpha_\mathrm{kick} \sqrt{2 E_\mathrm{kin} m_\mathrm{ej}}.
\end{equation}
In the early phase when the natal kick is determined, $E_\mathrm{kin}$ is of the order of the diagnostic explosion energy $E_\mathrm{expl}$ (i.e.\ the net energy of unbound material), within a factor of $2$--$3$ in recent 3D neutrino hydrodynamics simulations. Unlike \cite{janka2017}, we simply identify $E_\mathrm{kin}$ and $E_\mathrm{expl}$ so that we obtain the natal kick velocity $v_\mathrm{kick}$ as
\begin{equation}
\label{eq:vkick}
v_\mathrm{kick} = \frac{\alpha_\mathrm{kick} \sqrt{2 E_\mathrm{expl} m_\mathrm{ej}}}{m_\mathrm{NS}},
\end{equation}
where $m_\mathrm{NS}$ is the gravitational \ac{NS} mass. To obtain $m_\mathrm{ej}$, \cite{janka2017} related $E_\mathrm{expl}$ to the mass $m_\nu$ of the {\it neutrino--heated} ejecta via the nucleon recombination energy and then expressed $m_\mathrm{ej}$ as a multiple thereof.  The semi-analytical models of \cite{mueller2016simple} directly predict 
$m_\mathrm{ej}$,  $E_\mathrm{expl}$ and $m_\mathrm{NS}$ (see below), up to parameters based on 3D simulations and observational constraints.  These parameters are calibrated slightly differently than in \citet{mueller2016simple} (see Section \ref{sec:Muller}).  We can therefore work directly with Equation~(\ref{eq:vkick}).  

Equation~(\ref{eq:vkick}) needs to be evaluated at the time when the natal kick
asymptotes to its final value. One possibility, suggested by \cite{janka2017}, is to
relate the freeze-out of the natal kick to the termination of accretion
onto the \ac{NS}, which happens roughly when the post-shock velocity
equals the escape velocity \citep{marek2009,mueller2016simple}; this is
the criterion we adopt here.

Our key assumption is that the expectation value of the
anisotropy parameter $\alpha_\mathrm{kick}$ is independent of the
progenitor.  This is based on the observation that three-dimensional
explosion models \citep{melson2015b,lentz2015,mueller2017} with
multi-group neutrino transport typically develop unipolar or bipolar
explosions, i.e.\ there is limited variation in explosion
geometry. Moreover, there is a convergence to similar turbulent Mach
number around \citep{summa2016} and after shock revival, which implies
a similar density contrast between the under--dense neutrino-heated
bubbles and the surrounding down flows. This is somewhat dissimilar
from parameterised models \cite{wongwathanarat2013}, which show larger
variations in $\alpha_\mathrm{kick}$ because they can vary the explosion energy
independently of the progenitor structure.

While the assumption of uniform $\alpha_\mathrm{kick}$ is well motivated, some
caveats about its limitations are in order. Even though the
distribution of $\alpha_\mathrm{kick}$ may be relatively uniform across different
progenitors (which remains to be confirmed by more 3D explosion
models), $\alpha_\mathrm{kick}$ will show stochastic variations.  Moreover,
\ac{SN} models for progenitors with small CO cores are characterised
by medium-scale asymmetries \citep{wanajo2011,melson2015a} instead of
unipolar/bipolar modes during the explosion phase. 

Since theoretical arguments can only constrain the assumed uniform value of $\alpha_\mathrm{kick}$ within an order of magnitude, calibration is still required to roughly match the observed distribution of \ac{NS} natal kicks. The fit formulae presented below are based on a normalisation $\alpha_\mathrm{kick}=0.08$ that yields a match to the observed natal kick distribution of \citet{hobbs2005statistical}. 

\begin{figure*}
	\includegraphics[trim={4cm 2cm 4cm 2cm},clip,angle=90,width=\textwidth]{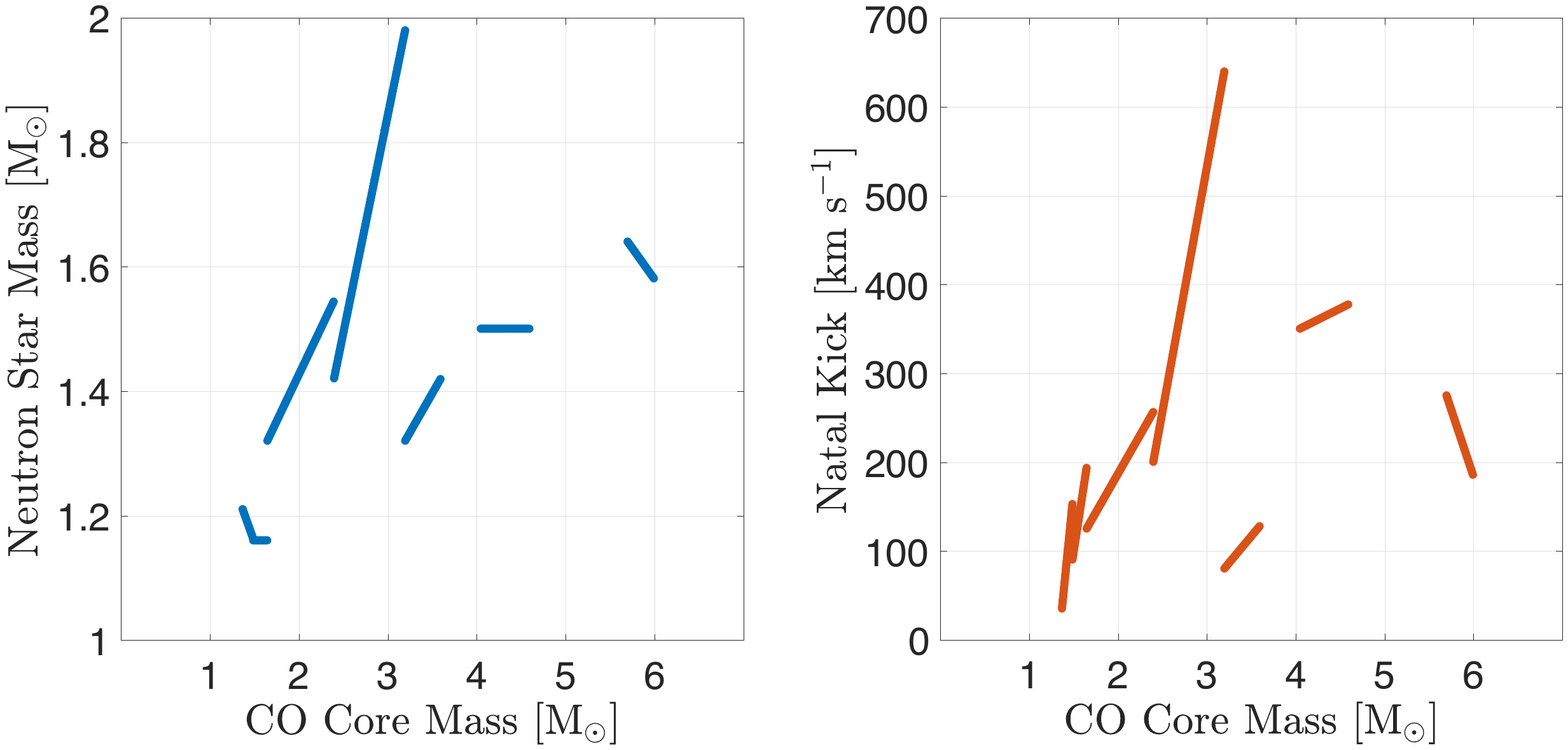}
\caption{
M\"uller \ac{SN} prescription of the best-fitting relation to the models described by \protect\cite{mueller2016simple} with parameters adjusted for better agreement with inferred \ac{SN} progenitor masses \citep{smartt2015}.  Gravitational mass (left) and natal kick (right) of the \ac{NS} as a function of the carbon--oxygen core mass. \ac{BH} formation is assumed to happen for $3.6\leq m_\mathrm{CO}<4.05$, $4.6\leq m_\mathrm{CO}<5.7$, and $m_\mathrm{CO}>6.0$, where $m_\mathrm{CO}$ is the 
carbon--oxygen core mass in $\rm M_{\rm \odot}$ units.
}
	\label{fig:Muller}
\end{figure*}

For the \ac{NS} mass $m_\mathrm{NS}$, we use 

\begin{equation}
\frac{m_\mathrm{NS}}{\rm M_{\odot}}
=
\left \{
\begin{array}{ll}
1.21-0.4  (m_\mathrm{CO}-1.372) ,& 1.372\leq  m_\mathrm{CO} < 1.49 \\
1.16,&  1.49\leq m_\mathrm{CO} < 1.65 \\
1.32 + 0.3 (m_\mathrm{CO}-1.65), & 1.65  \leq m_\mathrm{CO} < 2.4 \\
1.42 + 0.7 (m_\mathrm{CO}-2.4), & 2.4  \leq m_\mathrm{CO} < 3.2 \\
1.32+ 0.25 (m_\mathrm{CO}-3.2), & 3.2  \leq m_\mathrm{CO} < 3.6 \\
1.5  & 4.05  \leq m_\mathrm{CO}, < 4.6 \\
1.64 - 0.2 (m_\mathrm{CO}-5.7), & 5.7  \leq m_\mathrm{CO} < 6.
\end{array}
\right. ,
\end{equation}
where $m_\mathrm{CO}$ is the CO core mass in units of $\rm M_{\rm \odot}$. \ac{BH} formation is assumed to happen for
$3.6\leq m_\mathrm{CO}<4.05$,
$4.6\leq m_\mathrm{CO}<5.7$, and
$m_\mathrm{CO}>6.0$. 


The natal kicks are computed  as
\begin{equation}
\frac{v_\mathrm{kick}}{\mathrm{km}\, \mathrm{s}^{-1}}
=
\left\{
\begin{array}{ll}
	35+1000 (m_\mathrm{CO}-1.372) ,&  1.372  \leq m_\mathrm{CO} < 1.49 \\
	90 + 650 (m_\mathrm{CO}-1.49), & 1.49  \leq m_\mathrm{CO} < 1.65 \\
	100 + 175 (m_\mathrm{CO}-1.65), & 1.65  \leq m_\mathrm{CO} < 2.4 \\
	200 + 550 (m_\mathrm{CO}-2.4), & 2.4  \leq m_\mathrm{CO} < 3.2 \\
	80 + 120 (m_\mathrm{CO}-3.2), & 3.2  \leq m_\mathrm{CO} < 3.6 \\
	350 + 50 (m_\mathrm{CO}-4.05), & 4.05  \leq m_\mathrm{CO} < 4.6 \\
	275 - 300 (m_\mathrm{CO}-5.7), & 5.7  \leq m_\mathrm{CO} < 6.0
\end{array}
\right. .
\end{equation}

\section{Movement in the Galactic potential}
\label{sec:AppD}
\begin{figure*}
\vspace{-1in}
		\includegraphics[trim={2cm 0cm 3cm 0cm},clip,width=0.45\textwidth]{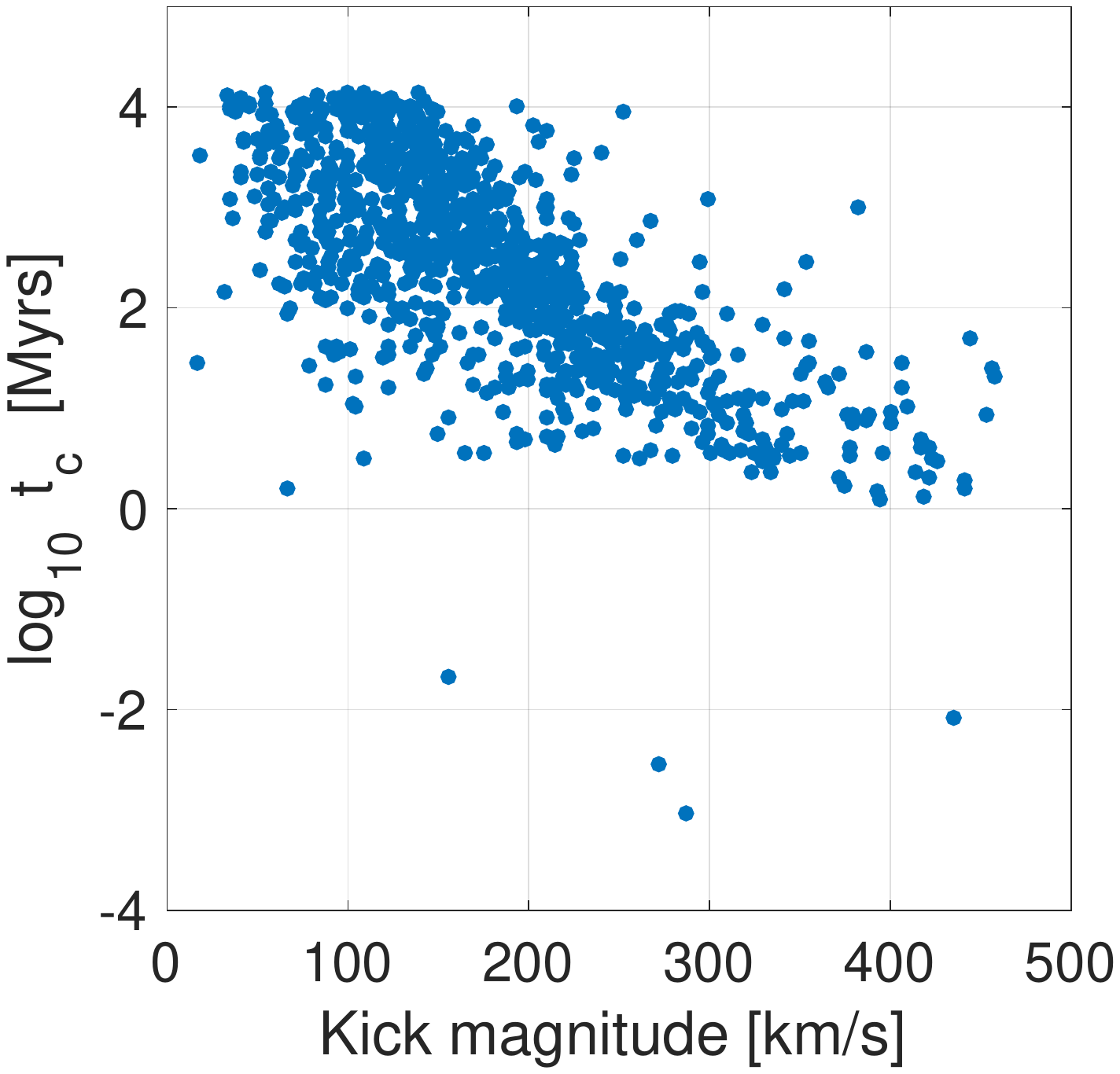}
		\includegraphics[trim={2cm 0cm 3cm 0cm},clip,width=0.45\textwidth]{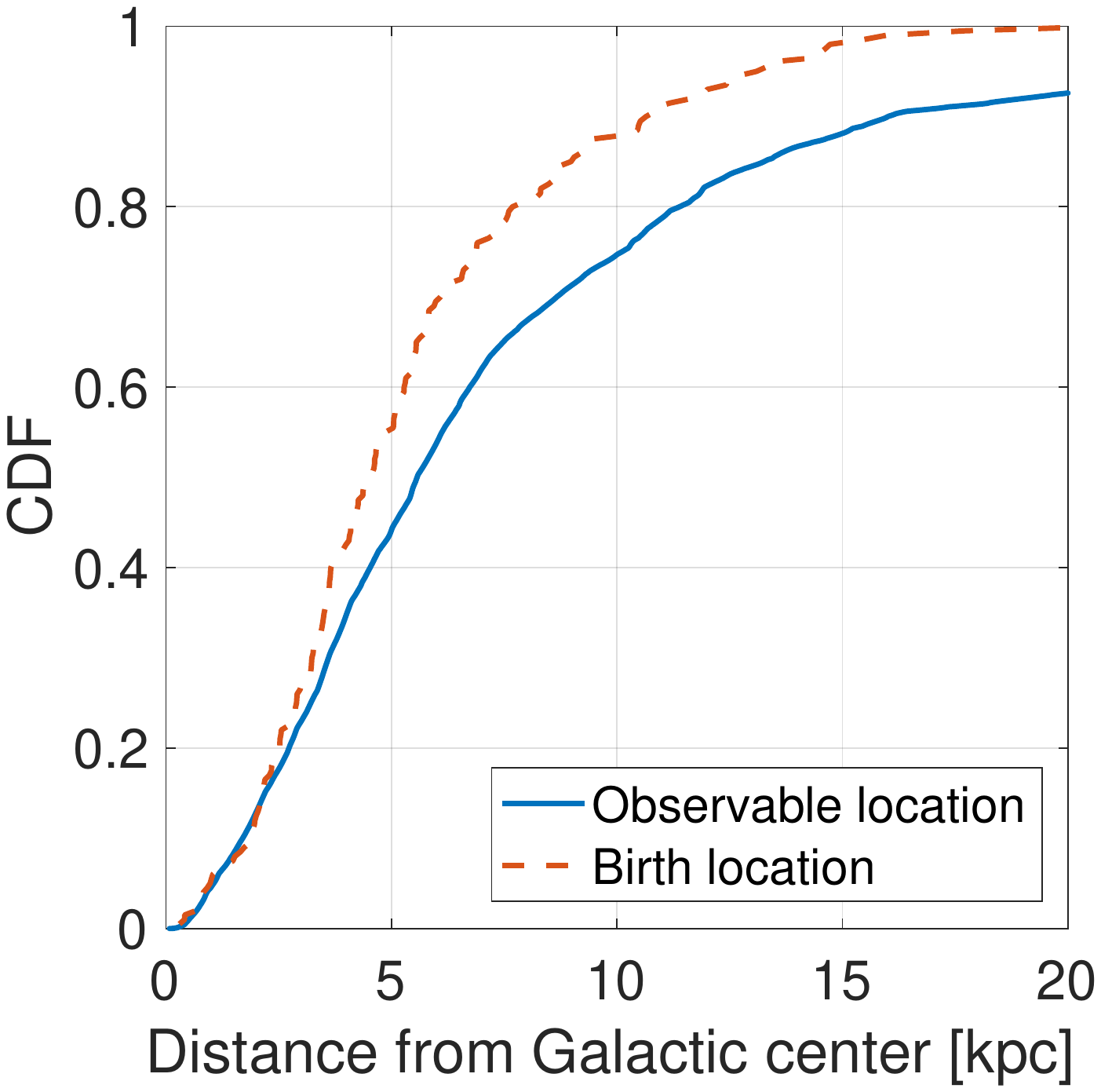}
\vspace{-1in}
		\caption{Scatter plot of the binary coalescence time against the DNS kicks magnitude in the \Fiducial~model (left panel).  DNS kicks are dominated by the Blaauw kick during the collapse of the secondary, which is proportional to the orbital velocity of the progenitor and therefore inversely correlated with the coalescence time of the binary.  These kicks spread the binaries in the Milky Way gravitational potential relative to birth sites, which are presumed to be in the disk plane (cumulative distribution function of the Galacto-centric distance for binaries born in the disk is shown in the right panel). 
		}
		\label{fig:kicks} 
\end{figure*}

\ac{DNS} centre-of-mass velocities in our \Fiducial~model, in which the second \ac{SN} is typically a \ac{USSN} with a low natal kick, are dominated by the Blaauw kick received as a result of the mass loss accompanying the collapse of the secondary.  This kick is proportional to the orbital velocity of the secondary before the collapse, which is greatest for the most compact binaries.  Therefore, the binary's velocity is anti-correlated with the coalescence time, as shown on the left panel of Figure~\ref{fig:kicks}.  If the \ac{USSN} progenitors are stripped even deeper than in COMPAS models during case~BB mass transfer \cite{tauris2015ultra}, as discussed in Section \ref{sec:discussion}, the mass lost during the \ac{SN} and the associated Blaauw kick would be further reduced.

These kicks have the effect of broadening the distribution of observed \ac{DNS} systems in the Galaxy.  We assume that each \ac{DNS} is formed in the thin disk, at $z=0$ in cylindrical coordinates, with a radial distribution proportional to the disk mass projected onto the Galactic equatorial plane.  We use model 2 of \citet{Irrgang:2013} 
for the Galactic matter distribution and total gravitational potential.  We do not account for scattering in this simplified analysis; while dynamical heating would increase the scale height of older populations, it does not appreciably impact the distribution of distances from the Galactic centre, which we estimate here.  After choosing a random initial location for the binary as above, we apply an additional initial velocity relative to the local rotational velocity with a magnitude equal to the binary's simulated kick velocity and a random direction.  The trajectory of the binary in the Galactic potential is solved with a Runge-Kutta integrator.  We sample the binary's subsequent motion at fixed time intervals between birth and merger (or a maximum age of 10 Gyr). The right panel of Figure~\ref{fig:kicks} shows the cumulative distribution function of the birth location, and the broader cumulative distribution function at which \ac{DNS} systems are expected to reside for a snapshot of all \acp{DNS} existing at the present moment.  The broadening of the distribution would be more significant in shallower gravitational potentials of less massive galaxies, which are probed with short \acp{GRB}.

\label{lastpage}
\end{document}